\documentclass[12pt]{article}
\usepackage{color,amssymb,epsfig,verbatim,amsmath}
\usepackage{amsfonts} 
\usepackage{longtable}
\usepackage{booktabs}
\usepackage{adjustbox}
\usepackage{accents}
\usepackage{caption}
\usepackage{pdflscape}
\usepackage{adjustbox}
\usepackage{tabularray}
\usepackage{accents}
\usepackage{array}
\usepackage{pdfpages}
\usepackage{tabu}
\usepackage[title]{appendix}
\def\dfrac#1#2{{\displaystyle{#1\over#2}}}

\def\boxit#1{\vbox{\hrule\hbox{\vrule\kern6pt
          \vbox{\kern6pt#1\kern6pt}\kern6pt\vrule}\hrule}}
\def\refhg{\hangindent=20pt\hangafter=1}
\def\refmark{\par\vskip 2mm\noindent\refhg}

\def\refhg{\hangindent=20pt\hangafter=1}
\def\refmark{\par\vskip 2mm\noindent\refhg}

\def\bse{\begin{eqnarray*}}
\def\ese{\end{eqnarray*}}
\def\be{\begin{eqnarray}}
\def\ee{\end{eqnarray}}
\def\bq{\begin{equation}}
\def\eq{\end{equation}}
\def\bse{\begin{eqnarray*}}
\def\ese{\end{eqnarray*}}

\usepackage{pgf,tikz}
\usetikzlibrary{arrows}
\usepackage{hyperref}
\hypersetup{
    colorlinks=true,
    linkcolor=blue,
    filecolor=magenta,      
    urlcolor=blue,
}
\usepackage{lscape}
\usepackage{pdflscape}
\usepackage{afterpage}

\newtheorem{theorem}{Theorem}[section]

\newtheorem{lemma}{Lemma}[section]

\makeatletter
\newcommand*\bigcdot{\mathpalette\bigcdot@{.5}}
\newcommand*\bigcdot@[2]{\mathbin{\vcenter{\hbox{\scalebox{#2}{$\m@th#1\bullet$}}}}}
\makeatother

\begin{document}
\thispagestyle{empty}

\hfill July 26, 2025 \\ \\

\baselineskip=28pt
\begin{center}
{\LARGE{\bf Causal Inference for Circular Data}}
\end{center}
\baselineskip=14pt
\vskip 2mm
\begin{center}
Kuan-Hsun Wu\footnote{Corresponding Author. Email: \hyperlink{mailto:110304015@g.nccu.edu.tw}{110304015@g.nccu.edu.tw}}
\\~\\

\textit{Department of Statistics, National Chengchi University}
\end{center}
\bigskip

\begin{center}
{\Large{\bf Abstract}}
\end{center}
\baselineskip=17pt
{
In causal inference, a fundamental task is to estimate the effect resulting from a specific treatment, which is often handled with inverse probability weighting. Despite an abundance of attention to the advancement of this task, most articles have focused on linear data rather than circular data, which are measured in angles. In this article, we extend the causal inference framework to accommodate circular data. Specifically, two new treatment effects, average direction treatment effect (ADTE) and average length treatment effect (ALTE), are introduced to offer a proper causal explanation for these data. As the average direction and average length describe the location and concentration of a random sample of circular data, the ADTE and ALTE measure the change in direction and length between two counterfactual outcomes. With inverse probability weighting, we propose estimators that exhibit ideal theoretical properties, which are validated by a simulation study. To illustrate the practical utility of our estimator, we analyze the effect of different job types on dispatchers' sleep patterns using data from Federal Railroad Administration.}

\par\vfill\noindent

\clearpage\pagebreak\newpage
\pagenumbering{arabic}

\newlength{\gnat}
\setlength{\gnat}{22pt}
\baselineskip=\gnat

\clearpage

\section{Introduction}
\textit{Circular statistics} deals with the data which can be regarded as points on the circumference of unit circle and has been widely applied in various fields. For instance, Nagasaki et al. (2025) modeled the traffic volume data within a day using mixtures of Kato-Jones distributions (Kato and Jones, 2015); Mardia et al. (2007) explored the protein structure using bivariate circular statistics; and Hansen and Mount (1990) utilized smooth techniques on data of \textit{stress field}, which describes how stress moves from point to point, and estimated the direction of stress. In these applications, data are typically observed rather than generated in a well-designed experiment. To correctly infer causality from such circular data, it is necessary to address the issue of confounding, which is a primary concern in the causal inference methodology.

Causal inference is the integration of methods exploring causality in observational studies. Specifically, it investigates the effect on an assigned outcome of interest brought by a treatment, which is usually characterized by the difference in two counterfactual outcomes. To broaden the general applicability of causal inference, a large body of literature has been contributed to accommodate data with complex features. For example, Yi and Chen (2023) and Chen (2020) handled the estimation of average treatment effect with measurement error. Since high dimensionality is common in the big data era, Huang and Chan (2017) focused on conditional average treatment effect where sufficient dimension reduction is taken into account. Moreover, Wu and Chen (2025) presented a flexible framework of estimating a diversity of casual effects where the dependence structure among confounders is described by graphical models. Despite abundant articles applying causal inference to complicated data, there is no special attention on the inferring causality from circular data.

In this article, we take the initiative in exploring causality for circular outcome. In particular, the location and dispersion of circular data are respectively measured by direction and length of its resultant vector. To capture the change in location and dispersion brought by a treatment, we apply inverse probability weighting (IPW) and define two new treatment effects: average length direction effect (ADTE) and 
average length treatment effect (ALTE). We derive the estimators for ADTE and ALTE under two IPW schemes, Horvitz and Thompson's (1952) version and H\'ajek's (1971) version. Moreover, we show that the estimators are consistent and asymptotically normal, which is further supported by a simulation study.

We arrange the article as follows. In Section \ref{Sec 2}, we introduce conditions for circular data and causal inference and define the ADTE and ALTE. The two IPW approaches to estimating them are explained in Section \ref{Sec 3}, in where the theoretical properties are subsequently given. We also study the estimators numerically and display the simulation results in Section \ref{Sec4}. In addition, we exemplify our estimators using Work Schedule and Sleep Patterns data of dispatchers in Section \ref{Sec 5}. Eventually, we summarize some significant results in Section \ref{Sec 6}.

\section{Treatment Effects for Circular Outcome}\label{Sec 2}
In this section, we introduce some notions of circular statistics and discuss the proper accommodation of circular data in causal inference.
\subsection{Circular Random Variables}
Let $\Theta$ be a continuous random variable with probability density function (PDF) $f(\theta)$ and cumulative distribution function (CDF) $F(\theta)$. We say $\Theta$ is a circular random variable if the following conditions are satisfied (e.g., Fisher, 1993):
\begin{itemize}
    \item[(C1)] $f(\theta)\ge0$.
    \item[(C2)] $f$ is $2\pi$-periodic. That is, $f(\theta) = f(\theta+2k\pi)$ for any integrer $k$.
    \item[(C3)] $\int_0^{2\pi}f(\theta)d\theta = 1$.
    \item[(C4)] $F(\theta) = \int_{0}^{\theta}f(t)dt$.
\end{itemize}
Condition (C1) assures that the PDF of $\Theta$ is non-negative, which is common even in an non-circular statistic context. Conditions (C2) and (C3) characterize the cyclic behavior of a circular random variable, which states the period of a cycle is of length $2\pi$. Lastly, although any $f$ meeting (C1)-(C3) are valid PDFs for circular random variables, it is convenient to regulate that the distribution starts at $0$, which is stated by the condition on CDF, (C4). Based on the target of analysis, one may change the starting point to other meaningful values.

A circular random variable is characterized by its $p$th trigonometric moment, which is defined as, for $p=1,2,\dots$,
\begin{equation}\label{pth trigo moment}
   \begin{aligned}
        \phi_p &\triangleq E\{\exp{(i\Theta)}\}\\
        &=\alpha_p + i \beta_p,
   \end{aligned}
\end{equation}
where 
\begin{equation}\label{pth alpha}
    \alpha_p \triangleq E(\cos{p\Theta})
\end{equation}
is the $p$th cosine moment of $\Theta$ and 
\begin{equation}\label{pth beta}
    \beta_p \triangleq E(\sin{p\Theta})
\end{equation}
is the $p$th sine moment of $\Theta$. Since (\ref{pth trigo moment}) can be viewed as a vector in the complex plane $\mathbb{C}$, we defined its length as 
\begin{equation}\label{pth length}
    \rho_p \triangleq (\alpha_p^2+\beta_p^2)^{1/2}
\end{equation}
and its direction as
\begin{equation}\label{pth direction}
    \begin{aligned}
        \mu_p &\triangleq \text{atan2}(\beta_p,\alpha_p)\\
        &=\begin{cases}
            \arctan(\beta_p/\alpha_p)\;\;&\text{if }\beta_p>0,\alpha_p>0; \\
             \arctan(\beta_p/\alpha_p)+\pi,\;\;&\text{if }\alpha_p<0;\\
              \arctan(\beta_p/\alpha_p)+2\pi,\;\;&\text{if }\beta_p>0,\alpha_p>0. \\
        \end{cases}
    \end{aligned}
\end{equation}
The quantities (\ref{pth length}) and (\ref{pth direction}) are called the $p$th {mean length} and {mean direction}, respectively. Since the case $p=1$ is frequently encountered, we simply write $\phi,\alpha,\beta,\rho$ and $\mu$ to refer to (\ref{pth trigo moment}), (\ref{pth alpha}), (\ref{pth beta}), (\ref{pth length}) and (\ref{pth direction}) with $p=1$, respectively.  In particular, we call $\phi$ the \textit{mean resultant vector} with \textit{mean resultant direction} $\mu\in[0,2\pi)$ describing the location at where the distribution centers and \textit{mean resultant length} $\rho\in[0,1]$ representing the extent of concentration. As $\arctan(\cdot)$ returns values on $[-\pi/2,\pi/2)$, rather than $[0,2\pi)$, a modified version of $\arctan(\cdot)$, atan2$(\cdot,\cdot)$, is used in (\ref{pth direction}) to return correct angles. 

Additionally, to perform meaningful analysis of circular data, we require the underlying distribution of the data has a well-defined $\mu$. Hence, the following condition is proposed:
\begin{itemize}
    \item[(C5)] Let $\Theta$ be a circular random variables satisfying (C1)-(C4). We have that $|\alpha|\in(0,1]$.
\end{itemize}

\subsection{Counterfactual Framework for Circular Outcomes}
 Let $A$ be a binary indicator of treatment membership, where $A=1$ suggests membership in the treated group and $A=0$ suggests membership in the control group, and let $\bold{X} = (1,X_1,\dots,X_p)$ be the $p$-dimensional covariates with intercept and its realization is denoted by $\bold{x}$. We denote the outcome of interest by $\Theta$. Let the observable random sample be $\mathcal{O} = (A_i,\bold{X}_i,\Theta_i)_{i=1}^n$. When $A=a$, the potential outcome under treatment condition $a$ is denoted by $\Theta^{(a)}$. Specifically, the observable outcome $\Theta$ and potential outcomes $\Theta^{(1)}$ and $\Theta^{(0)}$ follow circular distributions meeting conditions (C1)-(C5). Moreover, for $a\in\{0,1\}$, the first sine moment, first cosine moment, mean length and mean direction of $\Theta^{(a)}$ are denoted by $\alpha^{(a)},\beta^{(a)},\mu^{(a)}$ and $\rho^{(a)}$, respectively.

 We define the \textit{average direction treatment effect} (ADTE) as 
 \begin{equation}\label{ADTE}
     \begin{aligned}
         \tau &\triangleq \text{atan2}(\beta^{(1)},\alpha^{(1)}) - \text{atan2}(\beta^{(1)},\alpha^{(1)})\\
         &= \mu^{(1)} - \mu^{(0)},
     \end{aligned}
 \end{equation}
 which is the difference in the mean direction of two potential outcomes. Similarly, the \textit{average length treatment effect} (ALTE), which captures the change in mean length between two counterfactual outcomes, is given by
 \begin{equation}\label{ALTE}
     \begin{aligned}
         \xi &\triangleq \{(\alpha^{(1)})^2 +(\beta^{(1)})^2\}^{1/2} - \{(\alpha^{(0)})^2 +(\beta^{(0)})^2\}^{1/2}\\
         &= \rho^{(1)} - \rho^{(0)}.
     \end{aligned}
 \end{equation}

In this article, the issue is to estimate (\ref{ADTE}) and (\ref{ALTE}), quantities determined by the distribution $\Theta^{(1)}$ and $\Theta^{(0)}$, while the random sample $\mathcal{O}$ always lack either $\Theta^{(1)}$ or $\Theta^{(0)}$ for each instance. To solve this, we apply the inverse probability weighting (IPW; Rosenbaum and Rubin, 1983) approach. It primarily relies on the use of propensity score, which is given by
\begin{equation}\label{PS def}
    \pi(\bold{x}) \triangleq P(A=1\mid\bold{X} = \bold{x}).
\end{equation}
That is, the propensity score is the conditional probability of receiving treatment given $\bold{X} = \bold{x}$. To prepare for further developments in estimating (\ref{ADTE}) and (\ref{ALTE}), we introduce some common necessary assumptions for causal inference:
\begin{itemize}
\item[(C6)] {\it Consistency}: 

The observed outcome $\Theta$ is a mixture of the potential outcomes $(\Theta^{(1)},\Theta^{(0)})$. In other words, we have $\Theta = A\Theta^{(1)} + (1-A)\Theta^{(0)}$.
\item[(C7)] {\it Strong Ignorability}: 

Conditioning on $\bold{X}$, the treatment $A$ is independent of the joint distribution of $(\Theta^{(1)},\Theta^{(0)})$.

\item[(C8)] $0 < \pi \left(\boldsymbol{x} \right) < 1$.

\item[(C9)] $(A,\mathbf{X})$ is independent and identically distributed.

\item[(C10)] $\mathbf{X}$ is bounded.
\end{itemize}
With (C6) - (C10), numerous counterfactual quantities can be identified. The following lemma gives a broad class of identification results.
\begin{lemma}[Inverse Probability Weighting.]\label{IPW lemma}
    Let $\Theta, \Theta^{(1)}$ and $\Theta^{(0)}$ be circular random variables satisfying conditions (C1)-(C4) and assume that (C5) - (C9) hold. Suppose $g(\Theta)$ and $k(\bold{X})$ are functions depending only on $\Theta$ and $\bold{X}$, respectively. We have 
    \begin{equation}\label{ipw treated}
        E\bigg(\frac{A}{\pi(\bold{X})}g(\Theta)k(\bold{X})\bigg) = E\bigg(g(\Theta^{(1)})k(\bold{X})\bigg)
    \end{equation}
    and 
    \begin{equation}\label{ipw control}
    E\bigg(\frac{1-A}{1-\pi(\bold{X})}g(\Theta)k(\bold{X})\bigg) = E\bigg(g(\Theta^{(0)})k(\bold{X})\bigg).      
    \end{equation}
\end{lemma}
With $g(\Theta)$ specified $\cos{\Theta}$ or $\sin{\Theta}$ and $k(\bold{X}) =1$, Lemma \ref{IPW lemma} gives the identification of $\alpha^{(a)}$ and $\beta^{(a)}$ for $a=0,1$. Since the pair of estimands, ($\tau,\xi$), is a unique function of $(\alpha^{(1)},\beta^{(1)},\alpha^{(0)},\beta^{(0)})$, it is identifiable. In addition to identification, Lemma \ref{IPW lemma} is useful in the derivation of theoretical properties later shown in Section \ref{Sec 3.1}. The proof of Lemma \ref{IPW lemma} is placed in Appendix \ref{Lemma proof}.
\section{Methodology}\label{Sec 3}
\subsection{Inverse Probability Weighted Estimators}\label{Sec 3.1}
Motivated by Lemma \ref{IPW lemma}, we can construct consistent estimators based on the random sample $\mathcal{O}$. However, in reality, the propensity score $\pi(\cdot)$ is rarely known and should be estimated. In this study, we model the propensity score by logistic regression, which is given by
\begin{equation}\label{PS model}
    \pi(\bold{x}) = \frac{\exp{(\bold{x}}^\top\boldsymbol{\eta})}{1+\exp{(\bold{x}^\top\boldsymbol{\eta})}},
\end{equation}
where $\boldsymbol{\eta}\triangleq (\eta_0,\eta_1,\dots,\eta_p)^\top$. To estimate $\boldsymbol{\eta}$ in (\ref{PS model}), we apply maximum likelihood estimation, which yields the estimator $\widehat{\boldsymbol{\eta}}$ that solves the following estimating equation
\begin{equation}\label{logistic EE}
    \sum_{i=1}^n \psi_{\boldsymbol{\eta}} (A_i,\bold{X}_i;\boldsymbol{\eta})\triangleq \sum_{i=1}^n \{A_i - \pi(\bold{X}_i)\}\bold{X}_i^\top=0.
\end{equation}
Therefore, we can estimate (\ref{PS def}) with
\begin{equation}\label{PS estimate}
     \widehat{\pi}(\bold{x}) = \frac{\exp{(\bold{x}}^\top\widehat{\boldsymbol{\eta}})}{1+\exp{(\bold{x}^\top\widehat{\boldsymbol{\eta}})}}.
\end{equation}
Subsequently, we define the inverse probability weight for the treated outcome on the $i$th instance as 
\begin{equation}\label{HT weight 1}
    w_{i,1}^{(1)}\triangleq \frac{A_i}{n\widehat{\pi}(\bold{X}_i)}
\end{equation}
and that for the controlled outcome on the $i$th instance as 
\begin{equation}\label{HT weight 0}
    w_{i,1}^{(0)}\triangleq \frac{1-A_i}{n\{1-\widehat{\pi}(\bold{X}_i)\}},
\end{equation}
with $i=1,\dots,n$. Based on Lemma \ref{IPW lemma}, consistent estimators for $\alpha^{(a)}$ and $\beta^{(a)}$ are given by
\begin{equation}\label{HT cos}
    \widehat{\alpha}^{(a)} \triangleq \sum_{i=1}^n  w_{i,1}^{(a)}\cos{\Theta_i}
\end{equation}
and 
\begin{equation}\label{HT sin}
    \widehat{\beta}^{(a)} \triangleq \sum_{i=1}^n  w_{i,1}^{(a)}\sin{\Theta_i},
\end{equation}
for $a\in\{0,1\}$, respectively. To obtain the estimator of  (\ref{ADTE}) and (\ref{ALTE}), we directly plug in the estimators of $\alpha^{(a)}$ and $\beta^{(a)}$, which gives
\begin{equation}\label{HT ADTE}
    \widehat{\tau} \triangleq \text{atan2}(\widehat{\beta}^{(1)}, \widehat{\alpha}^{(1)}) -\text{atan2}(\widehat{\beta}^{(0)}, \widehat{\alpha}^{(0)}) 
\end{equation}
and
\begin{equation}\label{HT ALTE}
     \widehat{\xi} \triangleq [ \{\widehat{\alpha}^{(1)}\}^2 + \{\widehat{\beta}^{(1)}\}^2   ]^{1/2} -  [ \{\widehat{\alpha}^{(0)}\}^2 + \{\widehat{\beta}^{(0)}\}^2   ]^{1/2},
\end{equation}
respectively. Since the weights (\ref{HT weight 1}) and (\ref{HT weight 0}) may result in unstable estimators for extreme values of $\widehat{\pi}(\cdot)$, an alternative weight is given by normalizing them, which gives the following weight 
\begin{equation}\label{Hajek weight}
    w_{i,2}^{(a)} \triangleq\bigg(\sum_{j=1}^n  w_{j,1}^{(a)} \bigg)^{-1}  w_{i,1}^{(a)}
\end{equation}
for $a\in\{0,1\}$. Accordingly, we define the estimators of $\alpha^{(a)}$, $\beta^{(a)}$ using (\ref{Hajek weight}), which can be formulated as 
\begin{equation}\label{Hajek alpha}
    \widetilde{\alpha}^{(a)} \triangleq \sum_{i=1}^n w_{i,2}^{(a)} \cos{\Theta_i},
\end{equation}
\begin{equation}\label{Hajek beta}
    \widetilde{\beta}^{(a)} \triangleq \sum_{i=1}^n w_{i,2}^{(a)} \sin{\Theta_i},
\end{equation}
In a similar manner, (\ref{ADTE}) and (\ref{ALTE}) can be estimated with
\begin{equation}\label{Hajek ADTE}
    \widetilde{\tau} \triangleq \text{atan2}(\widetilde{\beta}^{(1)},\widetilde{\alpha}^{(1)}) -  \text{atan2}(\widetilde{\beta}^{(0)},\widetilde{\alpha}^{(0)})
\end{equation}
and
\begin{equation}\label{Hajek ALTE}
    \widetilde{\xi} \triangleq [ \{\widetilde{\alpha}^{(1)}\}^2 + \{\widetilde{\beta}^{(1)}\}^2   ]^{1/2} -  [ \{\widetilde{\alpha}^{(0)}\}^2 + \{\widetilde{\beta}^{(0)}\}^2   ]^{1/2},
\end{equation}
respectively. We call the weights (\ref{HT weight 1}) and (\ref{HT weight 0}) HT weights and the corresponding estimators (\ref{HT cos}), (\ref{HT sin}), (\ref{HT ADTE}) and (\ref{HT ALTE}) HT-type estimators since the use of this type of weight can be dated back to Horvitz and Thompson (1952). Likewise, the weight (\ref{Hajek weight}) can be attributed to H\'ajek (1971) and is referrd to as H\'ajek weight. The estimators (\ref{Hajek alpha}), (\ref{Hajek beta}), (\ref{Hajek ADTE}) and (\ref{Hajek ALTE}) which involves using  H\'ajek weight are called H\'ajek-type estimators.

We observe that two different weighting schemes lead to identical estimators of $\tau$ defined in (\ref{ADTE}), which is directly shown by
\begin{equation}\label{same ratio}
    \begin{aligned}
        \frac{\widetilde{\beta}^{(a)}}{\widetilde{\alpha}^{(a)}} &= \dfrac{\sum\limits_{i=1}^n w_{i,2}^{(a)}\cos{\Theta_i}}{\sum\limits_{i=1}^n w_{i,2}^{(a)}\sin{\Theta_i}}\\
        &= \dfrac{\sum\limits_{i=1}^n  \bigg(\sum\limits_{j=1}^n  w_{j,1}^{(a)} \bigg)^{-1}  w_{i,1}^{(a)}\cos{\Theta_i}}{\sum\limits_{i=1}^n  \bigg(\sum\limits_{j=1}^n  w_{j,1}^{(a)} \bigg)^{-1}  w_{i,1}^{(a)}\sin{\Theta_i}}\\
        &=\dfrac{\sum\limits_{i=1}^n w_{i,1}^{(a)}\cos{\Theta_i}}{\sum\limits_{i=1}^n w_{i,1}^{(a)}\sin{\Theta_i}}\\
        &=\frac{\widehat{\beta}^{(a)}}{\widehat{\alpha}^{(a)}}.
    \end{aligned}
\end{equation}
for $a\in\{0,1\}$. Since $\widehat{\tau}$ and $\widetilde{\tau}$ only depend on the ratios $\widehat{\beta}^{(a)}/\widehat{\alpha}^{(a)}$ and $\widetilde{\beta}^{(a)}/\widetilde{\alpha}^{(a)}$, for $a\in\{0,1\}$, respectively, (\ref{same ratio}) is sufficient for the result  $\widehat{\tau} = \widetilde{\tau}$. Hence, employing different weighting schemes only results in different estimator of $\xi$ in (\ref{ALTE}).
\subsection{Theoretical Results}
To clearly address the properties of IPW estimators, we introduce some notations. Let the vector of nuisance parameters be $\boldsymbol{\omega} = (\alpha^{(1)},\beta^{(1)},\alpha^{(0)},\beta^{(0)})^\top$ and the vector of parameters of interest be $\boldsymbol{\Delta}=(\tau,\xi)^\top$. The HT-type estimator of $\boldsymbol{\omega}$ and $\boldsymbol{\Delta}$ are denoted by $\widehat{\boldsymbol{\omega}} \triangleq (\widehat\alpha^{(1)},\widehat\beta^{(1)},\widehat\alpha^{(0)},\widehat\beta^{(0)})^\top$ and $\boldsymbol{\widehat{\Delta}}=(\widehat{\tau},\widehat{\xi})^\top$, respectively. Similarly, the H\'ajek-type estimator of $\boldsymbol{\omega}$ and $\boldsymbol{\Delta}$ are denoted by $\widetilde{\boldsymbol{\omega}} \triangleq (\widetilde\alpha^{(1)},\widetilde\beta^{(1)},\widetilde\alpha^{(0)},\widetilde\beta^{(0)})^\top$ and $\boldsymbol{\widetilde{\Delta}}=(\widetilde{\tau},\widetilde{\xi})^\top$, respectively. Given a positive integer $q$, we write the zero-valued column vector of length $q$ as $\bold{0}_q$. Additionally, the Fisher information matrix of $\boldsymbol{\eta}$ is denoted by $\mathcal{A}\triangleq E [\pi(\bold{X})\{1-\pi(\bold{X})\}\bold{X}^\top\bold{X}]$. We begin with stating the large sample property of nuisance parameters. 
\begin{theorem}\label{nuisance asymptotics} Under conditions (C1)-(C10), we have that, as $n\to\infty$,
    \begin{itemize}
        \item[(i)]$n^{1/2}(\widehat{\boldsymbol{\eta}}-\boldsymbol{\eta})\overset{d}{\longrightarrow}\textup{N}(\bold{0}_{1+p},\mathcal{A}^{-1})$;
        \item[(ii)] $n^{1/2}(\widehat{\boldsymbol{\omega}}-\boldsymbol{\omega})\overset{d}{\longrightarrow}\textup{N}(\bold{0}_{4},-\mathcal{B}\mathcal{A}^{-1}\mathcal{B}^\top+\mathcal{C})$;
        \item[(iii)] $n^{1/2}(\widetilde{\boldsymbol{\omega}}-\boldsymbol{\omega})\overset{d}{\longrightarrow}\textup{N}(\bold{0}_{4},-\mathfrak{B}\mathcal{A}^{-1}\mathfrak{B}^\top + \mathfrak{C})$,
    \end{itemize}
    where "$\overset{d}{\longrightarrow}$" denotes convergence in distribution and the components of variance-covariance matrix $\mathcal{B},\mathcal{C},\mathfrak{B}$ and $\mathfrak{C}$ are given in Appendix \ref{nuisance asymptotics proof}. 
\end{theorem}
Theorem \ref{nuisance asymptotics} tackles the consistency and asymptotic normality of IPW estimators of nuisance parameters under two weighting schemes. Firstly, (i) states that the coefficients of logistic regression are consistently estimated. Secondly, HT estimators in (ii) and H\'ajek estimators in (iii), which use the estimated propensity score involving (i), are consistent and asymptotically normal. With nuisance parameters handled by Theorem \ref{nuisance asymptotics}, we can obtain the estimators of $\boldsymbol{\Delta}$ by regarding it as a function of the  vector of nuisance parameter $\boldsymbol{\omega}$. Accordingly, we arrive at the following theorem by applying delta method.
\begin{theorem}\label{parameters of interest asymptotics}
    Under conditions (C1)-(C10), we have that, as $n\to\infty$,
    \begin{itemize}
        \item[(i)] $n^{1/2}(\widehat{\boldsymbol{\Delta}} - \boldsymbol{\Delta})\overset{d}{\longrightarrow}\textup{N}(\bold{0}_{2},\boldsymbol{\Sigma}_{\textup{HT}})$;
        \item[(ii)] $n^{1/2}(\widetilde{\boldsymbol{\Delta}} - \boldsymbol{\Delta})\overset{d}{\longrightarrow}\textup{N}(\bold{0}_{2},\boldsymbol{\Sigma}_{\textup{H\'ajek}})$,
    \end{itemize}
    where the variance-covariance matrices $\boldsymbol{\Sigma}_{\textup{HT}}$ and $\boldsymbol{\Sigma}_{\textup{H\'ajek}}$ and the entries of them are shown in Appendix \ref{parameters of interest proof}. 
\end{theorem}
With Theorem \ref{parameters of interest asymptotics}, we confirm that the proposed estimators form decent estimators of $\boldsymbol{\Delta}$ and can be used to estimate ADTE (\ref{ADTE}) and ALTE in (\ref{ALTE}).
\section{Numerical Studies}\label{Sec4}
\subsection{Simulation Design}
We explore the numerical behaviors of estimators (\ref{HT ADTE}), (\ref{HT ALTE}), (\ref{Hajek ADTE})  and (\ref{Hajek ALTE}).
Let the dimension of covariates be $p=3$. We consider three different sample sizes $n=250, 500$ or $1000$. The vector of covariate is generated as $\bold{X} = (1,X_1,X_2,X_3)^\top$, where $X_1,X_2$ and $X_3$ independently follow the Beta$(2,1)$ distribution.

\begin{figure}[h!
]
        \centering
        \includegraphics{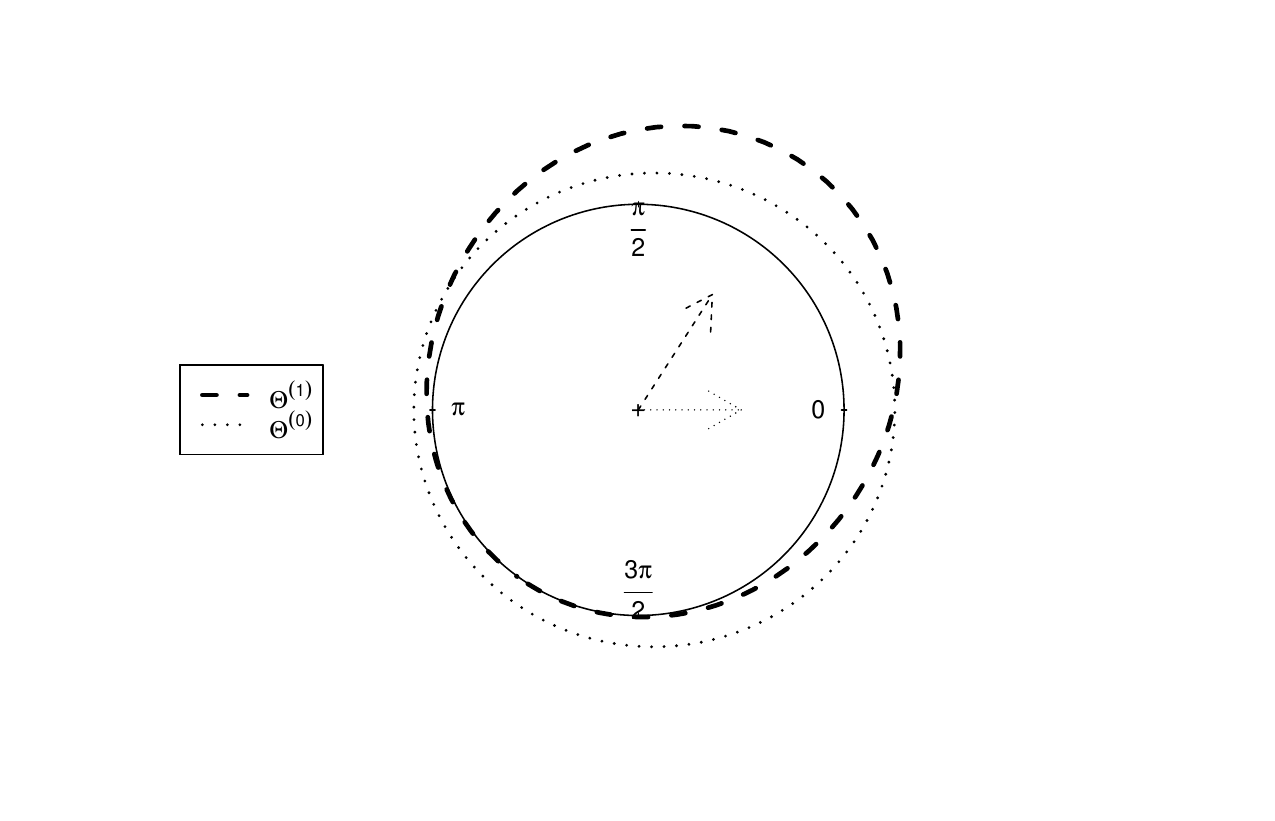}
        \caption{Illustration for simulation design. The ADTE and ALTE are the differences in the angles and lengths between two resultant vectors, respectively. The potential outcomes $\Theta^{(1)}$ and $\Theta^{(0)}$ have mean direction $1$ and $0$ and mean length $4/6$ and $1/2$, respectively.}
        \label{Y1Y0_circular}
    \end{figure}

When $\mathbf{X}$ is generated, the treatment $A$ is generated by the following model
\begin{equation}\label{model A}
\text{logit}\left\{ P(A=1|\mathbf{X}) \right\} = 1 + X_1 +X_2+ X_3.
\end{equation}
When generating potential outcomes, we consider four different scenarios where either ADTE or ALTE is subject to the  confounders $\bold{X}$. Let WC$(\mu,\rho)$ denote the wrapped Cauchy distribution with parameters mean direction $\mu$ and mean resultant length $\rho$. The details of the parameter specification is described as follows.
\begin{itemize}
    \item[(i)] Scenario 1: Heterogeneous ADTE and constant ALTE. 
    $$
 \Theta^{(1)} \mid\bold{X}\sim \textup{WC}\bigg(X_1+X_2+X_3,\frac{5}{6}\bigg); 
 \Theta^{(0)} \mid\bold{X} \sim \textup{WC}\bigg(\frac{1}{2}(X_1+X_2+X_3),\frac{2}{3}\bigg)
    $$
    \item[(ii)] Scenario 2: Constant ADTE and heterogeneous ALTE.
    $$
        \Theta^{(1)} \mid\bold{X}\sim \textup{WC}\bigg(1,\frac{1}{3}(X_1+X_2+X_3)\bigg);    \Theta^{(0)} \mid\bold{X} \sim \textup{WC}\bigg(0,\frac{1}{4}(X_1+X_2+X_3)\bigg)
    $$
    \item[(iii)] Scenario 3: Heterogeneous ADTE and heterogeneous ALTE.
    $$
\begin{aligned}
     &\Theta^{(1)} \mid\bold{X}\sim \textup{WC}\bigg(X_1+X_2+X_3,\frac{1}{3}(X_1+X_2+X_3)\bigg);    \\
 &\Theta^{(0)} \mid\bold{X} \sim \textup{WC}\bigg(\frac{1}{2}(X_1+X_2+X_3),\frac{1}{4}(X_1+X_2+X_3)\bigg)
\end{aligned}
    $$
\end{itemize}
Finally, the observed random variable $\Theta_i$ is acquired by the consistency assumption, which gives
\begin{equation}
    \Theta_i = A\Theta^{(1)}_i +(1-A)\Theta^{(0)}_i
\end{equation}
for $i=1,\dots,n$. Consequently, we obtain the IID sample $\mathcal{O}$ of size $n$. In all scenarios, the true ADTE and ALTE are set to $1$ and $1/6$, respectively. For demonstration purpose, Figure \ref{Y1Y0_circular} illustrates the potential outcomes with different circular distribution.

For each setting,  we generate $1000$ datasets and obtain the estimate of (\ref{HT ADTE}), (\ref{HT ALTE}), (\ref{Hajek ADTE}) and (\ref{Hajek ALTE}). To examine the performance of the estimators derived by the proposed methods, we compute several indexes including the biases (BIAS), standard errors (S.E.), mean squared errors (MSE) and coverage rate (CR) using the produced $1000$ estimates. 
\subsection{Simulation Result}
Table \ref{sim table} collects the simulation results across all four scenarios. Overall, the numerical results offer strong empirical evidence that the proposed estimators can consistently estimate the parameter, regardless of heterogeneity. Moreover, since the coverage rates for all cases are close to $95\%$, the asymptotic normality shown in Theorem \ref{parameters of interest asymptotics} is verified numerically.

For $\tau$ defined in (\ref{ADTE}), both the HT and H\'ajek estimators shown in  (\ref{HT ADTE}) and (\ref{Hajek ADTE}) yield identical point estimates, which aligns with the theoretical finding established in (\ref{same ratio}). Across scenarios with either constant or heterogeneous directional treatment effects, the estimators of (\ref{ADTE}) exhibit decreasing bias and MSE as the sample size increases. As for $\xi$ in (\ref{ALTE}), we observe that proposed estimators can offer precise estimates for all scenarios. Furthermore, the HT estimator (\ref{HT ALTE}) and the H\'ajek-type estimator (\ref{Hajek ALTE}) demonstrate opposite characteristics in small sample sizes. Although the HT-type estimator (\ref{HT ALTE}) tends to yield less biased estimates, it exhibits a larger variance. Conversely, the H\'ajek-type estimator (\ref{Hajek ALTE}) provides estimates with more stability, yet they are relatively more biased. To summarize, the simulation results confirm the equivalence of the HT and H\'ajek estimators for (\ref{ADTE}) and other large sample properties established in Theorem \ref{parameters of interest asymptotics}.

In addition, we comment that the characteristics of two types of estimators of (\ref{ALTE}) exhibit connections with the average treatment effect literature. Firstly, the H\'ajek estimator (\ref{Hajek ALTE}) generally shows greater stability than the HT-type estimator (\ref{HT ALTE}), which is in accordance with the numerical studies of Lunceford and Davidian (2004). Secondly, when estimating (\ref{ALTE}) in a small sample size, choosing one estimator over the other is a bias-variance tradeoff, which is a relationship stated in Khan and Ugander (2023). However, when the sample size $n$ is sufficiently large, the performance of the HT-type estimator (\ref{HT ALTE}) resembles that of the H\'ajek-type estimator (\ref{Hajek ALTE}), and the bias-variance tradeoff becomes negligible. 

\section{Data Analysis}\label{Sec 5}
In this section, we focus on analyzing the Work Schedules and Sleep Pattern Survey Data of Railroad Dispatchers (Federal Railroad Administration, 2008) with the proposed methods in Section \ref{Sec 3}. Each participant contributes his/her demographic information and a daily record of sleep schedule for the study. For comprehensive description of the survey methodology, one can visit the website \url{https://railroads.dot.gov/program-areas/human-factors/work-schedules-and-sleep-pattern-survey-data} for official information and the access of full dataset.

After necessary pre-processing, there are $439$ participants at total. The dataset contains a variety of variables, including age (\texttt{Age\_Group}, $1$ as 20-29 years old, $2$ as 30-39 years old, $3$ as 40-49 years old,$4$ as 50-59 years old and $6$ as older than 59), sex (\texttt{Sex}, $1$ as male and $2$ as female), the count of total years in the present job (\texttt{Total\_years\_present\_job}, in years), job types of dispatchers (\texttt{Job\_type}), self-rated health status (\texttt{Health\_Status}, $1$ as poor, $2$ as fair, $3$ as good and $4$ as excellent), marital status (\texttt{Marital\_Status}, $1$ as single and $2$ as married), whether or not ingesting caffeinated drink (\texttt{Caff\_Beverages}, $1$ as yes and $2$ as no), the time of falling asleep (\texttt{Time\_fell\_asleep}, in 24-hour clock form HH:MM with HH and MM being the hour and minute), self-rated frequency of feeling mentally or physically drained after work (\texttt{Mentally\_Drained} and \texttt{Phys\_Drained}, $1$ as never, $2$ as occasionally, $3$ as frequently and $4$ as always) and the count of life events in the last six months (\texttt{Total\_life\_events}), including birth of child, death of spouse, martial or financial woes, and other events.

We are interested in the effect of different job types of dispatchers on the time of falling asleep. In this dataset, there are two types of dispatchers: trick dispatchers and assistant chief dispatcher. The former regulates access to a designated territory and the latter is responsible of supervising and assisting the former. Moreover, we express the time of falling asleep as radians using the relationship $2\pi$ radians $=24$ hours. Using the notation in Section \ref{Sec 2}, the treatment variable $A$ is set as \texttt{Job\_type} with $1$ being assistant chief dispatcher and $0$ being the trick dispatcher and the outcome variable $\Theta$ is set as \texttt{Time\_fell\_asleep}, which is transformed as radians. Other mentioned variables which are not selected as treatment or outcome are considered confounders $\bold{X}$. 

\begin{landscape}
    \begin{figure}[h!]
    \centering
    \includegraphics[width=0.85\linewidth]{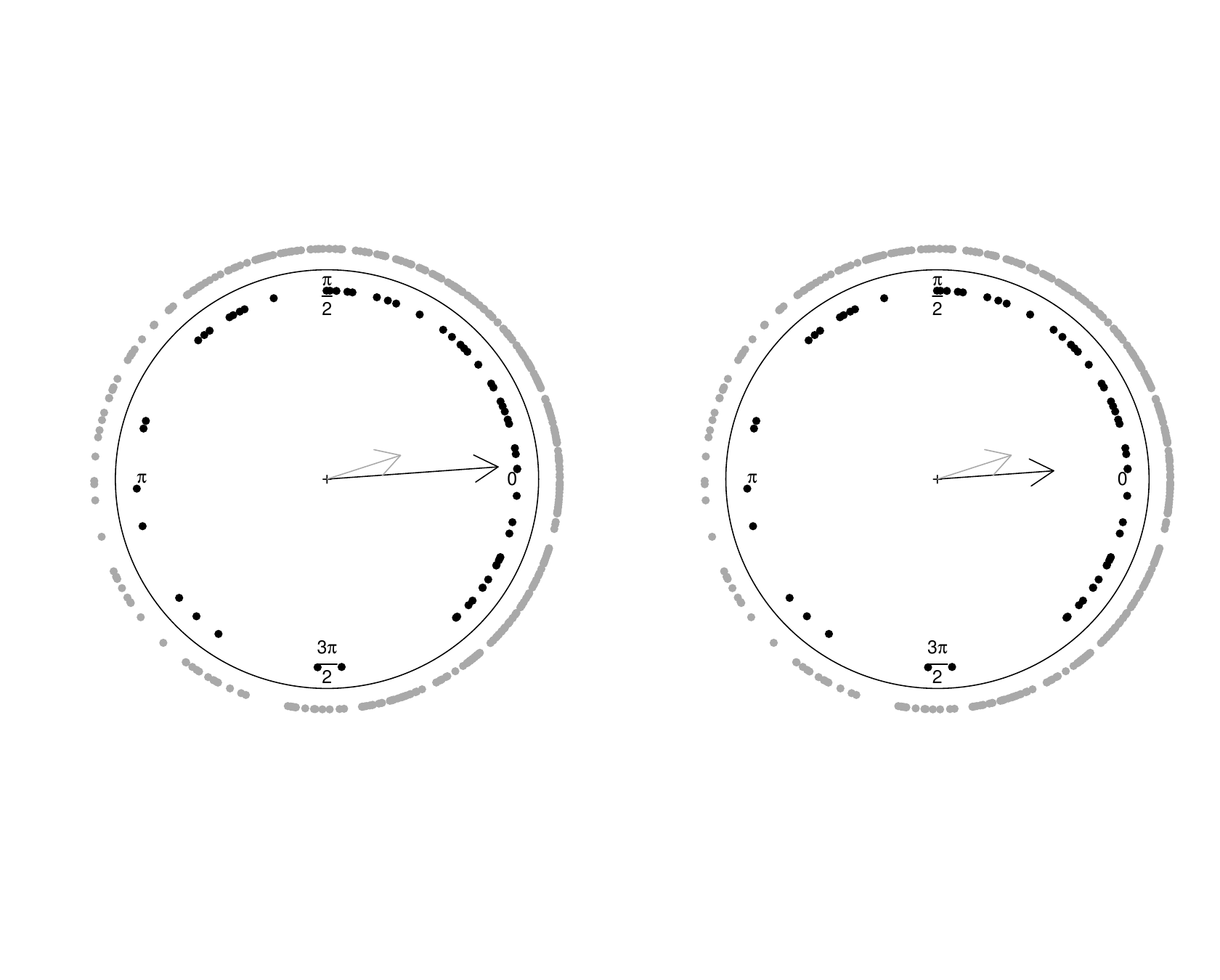}
    \caption{Left panel: the estimated resultant vectors of two counterfactual populations using HT weighting. Right panel: the estimated resultant vectors of two counterfactual populations using H\'ajek weighting. The black dots and arrows represent the participants with sleeping disorder and the gray dots and arrows represent those without disorder.}
    \label{data_analysis}
\end{figure}
\end{landscape}

Our goal is to estimate ADTE defined in (\ref{ADTE}) and ALTE defined in (\ref{ALTE}). The estimated ADTE using (\ref{HT ADTE}) and (\ref{Hajek ADTE}) is $-0.243$ radians, or equivalently, $-55.691$ minutes. Please note that we adopt the polar coordinate system in this article, which is counterclockwise, whereas the conventional 24-hour clock is clockwise. In this context, a negative ADTE estimate indicates that the treated group tends to fall asleep later. For ALTE, the HT-type estimator (\ref{HT ALTE}) yields an estimate of $0.445$ while the H\'ajek type estimator gives an estimates $0.184$. Since both estimators produce positive values, this supports that the assistant chief dispatchers exhibit more similarity in their falling-asleep times. In conclusion, the estimators suggests that the assistant chief dispatchers fall asleep approximately one hour later than trick dispatchers on average, and the falling-asleep times among the trick dispatchers are more dispersed than those among assistant chief dispatchers. For graphical illustration, the plots of the estimated resultant vector of two counterfactual distributions are placed at Figure \ref{data_analysis}.
\section{Summary}\label{Sec 6}
 In this article, we initiate to integrate circular statistics into the counterfactual framework. In particular, we introduce the ADTE and ALTE as population causal measures for circular data and, consequently, propose the corresponding estimators under HT and H\'ajek-type of inverse probability weighting. From the theoretical perspective, the proposed estimators enjoy several desirable properties such as consistency and asymptotic normality, which are further verified by the simulation study.

This article broadens the scope of causal inference by clarifying new types of causal effects for circular data as outcome variable. The contribution can be viewed as a critical advance and extension in both methodology and application. The proposed method is an extension of the commonly-used counterfactual framework and allows analysts to flexibly investigate causality in time, direction, or other types of data which can be placed on a unit circle.
\section*{Software}
The R code for simulation is available at \href{https://github.com/kuanhsun/Causal-Inference-for-Circular-Data}{https://github.com/kuanhsun/Causal-Inference-for-Circular-Data}.
\section*{Acknowledgement}
The author has notified FRA for use of the Work schedules and sleep patterns of railroad dispatchers data via e-mail and appreciated Dr. Li-Pang Chen for illuminating discussions.
\section*{References}
\refmark Chen, L. P. (2020). Causal inference for left-truncated and right-censored data with covariate measurement error. \textit{Computational and Applied Mathematics}, 39, 1-27.%

\refmark{Federal Railroad Administration. (2008).
\textit{Data Files: Work schedules and sleep patterns of railroad dispatchers}. Washington, DC: U.S. Department of Transportation.}

\refmark{Fisher, N. I. (1993). \textit{Statistical Analysis of Circular Data}. Cambridge University Press, Cambridge.}

\refmark{H\'ajek, J. (1971). Comment on "An Essay on the
Logical Foundations of Survey Sampling, Part
One," in \textit{The Foundations of Survey Sampling},
Godambe, V.P. and Sprott, D.A. eds., 236, Holt,
Rinehart, and Winston. }

\refmark{Hansen, K. M. and Mount, V. S. (1990). Smoothing and extrapolation of crustal stress orientation measurements. \textit{Journal of Geophysical Research: Solid Earth}, 95(B2), 1155-1165.}

\refmark{Horvitz, D. G. and Thompson, D. J. (1952). A generalization of sampling without replacement from a finite universe. \textit{Journal of the American statistical Association}, 47(260), 663-685.}

\refmark{Huang, M. Y. and Chan, K. C. G. (2017). Joint sufficient dimension reduction and estimation of conditional and average treatment effects. \textit{Biometrika}, 104(3), 583-596.}

\refmark{Kato, S. and Jones, M. C. (2015). A tractable and interpretable four-parameter family of unimodal distributions on the circle. \textit{Biometrika}, 102(1), 181-190.}

\refmark Khan, S. and Ugander, J. (2023). Adaptive normalization for IPW estimation. \textit{Journal of Causal Inference}, 11, 20220019.

\refmark Lunceford, J. K. and Davidian, M. (2004). Stratification and weighting via the propensity score in estimation of causal treatment effects: a comparative study. {\em Statistics in Medicine}, 23, 2937-2960 %

\refmark{Nagasaki, K., Kato, S., Nakanishi, W. and Jones, M. C. (2025). Traffic count data analysis using mixtures of Kato–Jones distributions. \textit{Journal of the Royal Statistical Society Series C: Applied Statistics}, 74(2), 352-372.}

\refmark{Mardia, K. V., Taylor, C. C. and Subramaniam, G. K. (2007). Protein bioinformatics and mixtures of bivariate von Mises distributions for angular data. \textit{Biometrics}, 63(2), 505-512.}

\refmark Rosenbaum, P. R. and Rubin, D. B. (1983) The central role of the propensity score in observational studies for causal effects. {\em Biometrika}, {70}, 41-55. %

\refmark{Wu, K. H. and Chen, L. P. (2025). A unified approach for estimating various treatment effects in causal inference. arXiv preprint \hyperlink{https://arxiv.org/abs/2503.22616}{arXiv:2503.22616}.}

\refmark Yi, G. Y. and Chen, L. P. (2023). Estimation of the average treatment effect with variable selection and measurement error simultaneously addressed for potential confounders. \textit{Statistical Methods in Medical Research}, 32, 691-711.%

\clearpage
\begin{landscape}
\captionsetup{width=18cm}
    \begin{longtable}{cclcccccccccccc}
    \caption{Simulation results of ADTE and ALTE under Scenario 1-3.}\label{sim table}\\
    \hline
    &&&\multicolumn{4}{c}{Scenario 1}&\multicolumn{4}{c}{Scenario 2}&\multicolumn{4}{c}{Scenario 3}\\
    \cmidrule(lr){4-7}\cmidrule(lr){8-11}\cmidrule(lr){12-15}
    $n$ & Estimand & Method & BIAS & S.E. & MSE & CR& BIAS & S.E. & MSE & CR& BIAS & S.E. & MSE & CR\\
   $250$ & $\tau$ & HT& 0.006& 0.369& 0.136  &0.949 & -0.014 &0.555& 0.308 & 0.935 &0.028& 0.603& 0.364   & 0.956\\
   && H\'ajek & 0.006& 0.369& 0.136  &0.949 & -0.014 &0.555& 0.308 & 0.935 &0.028& 0.603& 0.364   & 0.956\\
   &$\xi$&HT& -0.028& 0.177& 0.032   & 0.947&-0.056& 0.210 &0.047  &  0.941&-0.061& 0.216& 0.050   & 0.938 \\
   &&H\'ajek &-0.036 &0.161& 0.027  &0.969& -0.059& 0.192 &0.040 &0.954&-0.065& 0.197& 0.043 & 0.956 \\
   \hline
   $500$&$\tau$&HT &-0.005 &0.288 &0.083&0.952& 0.019& 0.383 &0.147&0.957&-0.016& 0.414& 0.172& 0.950 \\
   &&H\'ajek &-0.005 &0.288 &0.083&0.952& 0.019& 0.383 &0.147&0.957&-0.016& 0.414& 0.172& 0.950 \\
   &$\xi$&HT& -0.008 &0.119& 0.014& 0.947&-0.030& 0.141 &0.021& 0.950  &-0.018 &0.140 &0.020&0.947\\
   && H\'ajek& -0.010& 0.117 &0.014 &0.960& -0.031& 0.140 &0.020 &0.952& -0.020& 0.137& 0.019 &0.951\\
   \hline
   $1000$&$\tau$&HT&-0.006& 0.245& 0.060&0.952 &0.000 &0.230& 0.053    &0.951 &-0.002 &0.315 &0.099 & 0.956\\
   && H\'ajek &-0.006& 0.245& 0.060&0.952 &0.000 &0.230& 0.053    &0.951 &-0.002 &0.315 &0.099 & 0.956\\
   &$\xi$&HT&-0.005& 0.087 &0.008& 0.948 &-0.016& 0.103 &0.011   & 0.934& -0.011 &0.104& 0.011  &  0.945\\
   && H\'ajek& -0.005& 0.085& 0.007 &  0.956 &-0.016& 0.103 &0.011 & 0.939 &-0.011& 0.103 &0.011& 0.947\\
   \hline
\end{longtable}
\end{landscape}

\clearpage

\clearpage
\clearpage
\thispagestyle{empty}
\setcounter{footnote}{0}
\setcounter{section}{0}
\numberwithin{equation}{section}
\renewcommand{\thesection}{\alph{section}}

\hfill July 26, 2025\ \\

\baselineskip=28pt
\begin{center}
{\LARGE{\bf Supporting Information for “Causal Inference for Circular Data ”}}
\end{center}
\baselineskip=14pt
\vskip 2mm
\begin{center}
Kuan-Hsun Wu\footnote{Corresponding Author. Email: \hyperlink{mailto:110304015@g.nccu.edu.tw}{110304015@g.nccu.edu.tw}}
\\~\\

\textit{Department of Statistics, National Chengchi University}
\end{center}
\bigskip

\begin{center}
{\Large{\bf Abstract}}
\end{center}
\baselineskip=17pt
{

The supporting information contains the details of the lemma and proofs of theorems in the manuscript, entitled “Causal Inference for Circular Data”.
}

\par\vfill\noindent
\underline{\bf Keywords}: Causal inference; circular statistics; directional data; mean direction; mean resultant length;  inverse probability weighting.

\clearpage\pagebreak\newpage
\pagenumbering{arabic}

\setlength{\gnat}{22pt}
\baselineskip=\gnat

\clearpage
\setcounter{footnote}{0}
\setcounter{section}{0}
\appendix

\section{Proof of Lemma \ref{IPW lemma}}\label{Lemma proof}
To show (\ref{ipw treated}), we begin with observing that
\begin{equation}\label{ipw result}
    \begin{aligned}
    E\bigg(Ag(\Theta)\mid\bold{X}\bigg) &= E\bigg(1\times g(\Theta)\mid\bold{X},A=1\bigg) P(A=1\mid\bold{X})\\
    &\;\;\;+E\bigg(0\times g(\Theta)\mid\bold{X},A=0\bigg) P(A=0\mid\bold{X})\\
    &=E\bigg( g(A\Theta^{(1)}+(1-A)\Theta^{(0)})\mid\bold{X},A=1\bigg)\pi(\bold{X})\\
    &=E\bigg( g(\Theta^{(1)})\mid\bold{X},A=1\bigg)\pi(\bold{X})\\
    &=
    E\bigg( g(\Theta^{(1)})\mid\bold{X}\bigg)\pi(\bold{X}),
\end{aligned}
\end{equation}
where the first equality is obtained by partitioning the expectation with events $A=1$ and $A=0$, the second and third equality are applications of the consistency assumption (C5) and the fourth equality utilizes condition (C6). Moreover, applying the law of iterated expectations to the left-hand side of (\ref{ipw treated}) yields
\begin{equation} 
    \begin{aligned}
    E\bigg(\frac{A}{\pi(\bold{X})}g(\Theta)k(\boldsymbol{X})\bigg) &= E\Bigg\{\frac{k(\bold{X})}{\pi(\bold{X})}E\bigg(Ag(\Theta)\mid\bold{X}\bigg)\Bigg\} \\
    &=E\Bigg\{\frac{k(\bold{X})}{\pi(\bold{X})}E\bigg( g(\Theta^{(1)})\mid\bold{X}\bigg)\pi(\bold{X})\Bigg\}\\
    &=E\Bigg\{{k(\bold{X})}E\bigg( g(\Theta^{(1)})\mid\bold{X}\bigg)\Bigg\}\\
    &=E\Bigg\{E\bigg(g(\Theta^{(1)}){k(\bold{X})} \mid\bold{X}\bigg)\Bigg\}\\
    &=E\bigg(g(\Theta^{(1)}){k(\bold{X})} \bigg),
\end{aligned}
\end{equation}
where the second equality utilized (\ref{ipw result}). Hence, (\ref{ipw treated}) follows. With $A/\pi(\bold{X})$ replaced by $(1-A)/\{1-\pi(\bold{X})\}$, (\ref{ipw control}) can be proved in a similar manner. This gives Lemma \ref{IPW lemma}$.\hfill\square$
\section{Proof of Theorem \ref{nuisance asymptotics}}\label{nuisance asymptotics proof}
In this section, we prove the statement regarding the limiting distribution of estimators $\widehat{\boldsymbol{\omega}}$ and $\widetilde{\boldsymbol{\omega}}$ of the nuisance parameter $\boldsymbol{\omega}$ using M-estimation (Stefanski and Boos, 2002). The asymptotic results for the HT-type and H\'ajek-type estimators are handled separately. Since $\widehat{\boldsymbol{\omega}}$ and $\widetilde{\boldsymbol{\omega}}$ involve using $\widehat{\boldsymbol{\eta}}$ as a part of estimate, we collectively tackle their asymptotic distribution. We define the true parameter as $\boldsymbol{\vartheta}\triangleq (\boldsymbol{\eta}^\top,\boldsymbol{\omega}^\top)^\top$ with its HT estimator defined as $\widehat{\boldsymbol{\vartheta}}\triangleq(\widehat{\boldsymbol{\eta}}^\top,\widehat{\boldsymbol{\omega}}^\top)^\top$ and its H\'ajek estimator defined as $\widetilde{\boldsymbol{\vartheta}}\triangleq (\widehat{\boldsymbol{\eta}}^\top,\widetilde{\boldsymbol{\omega}}^\top)^\top$.\\
\textit{\underline{Part 1: HT-type estimators.}}\\
We represent the estimators in the form of estimating equations.  The estimating equation for the coefficients of logistic regression is readily presented in (\ref{logistic EE}). As for $\alpha^{(1)},\beta^{(1)},\alpha^{(0)}$ and $\beta^{(0)}$, the HT-type estimators in (\ref{HT cos}) and (\ref{HT sin}) with $a\in\{0,1\}$ are the solutions to
\begin{align}
     \sum_{i=1}^n \psi_{\alpha^{(1)}}^{\text{HT}}(A_i,\bold{X}_i,\Theta_i;\boldsymbol{\eta},\alpha^{(1)}) &\triangleq \sum_{i=1}^n \bigg(\frac{A_i}{\pi(\bold{X}_i)}\cos{\Theta}_i - \alpha^{(1)}\bigg)=0, \label{alpha 1 EE}\\
     \sum_{i=1}^n \psi_{\beta^{(1)}}^{\text{HT}} (A_i,\bold{X}_i,\Theta_i;\boldsymbol{\eta},\beta^{(1)})&\triangleq \sum_{i=1}^n \bigg(\frac{A_i}{\pi(\bold{X}_i)}\sin{\Theta}_i - \beta^{(1)} \bigg)=0,\label{beta 1 EE}\\
      \sum_{i=1}^n \psi_{\alpha^{(0)}}^{\text{HT}}(A_i,\bold{X}_i,\Theta_i;\boldsymbol{\eta},\alpha^{(0)}) &\triangleq \sum_{i=1}^n \bigg(\frac{1-A_i}{1-\pi(\bold{X}_i)}\cos{\Theta}_i - \alpha^{(0)}\bigg)=0\label{alpha 0 EE}
\end{align}
and
\begin{equation}\label{beta 0 EE}
    \sum_{i=1}^n \psi_{\beta^{(a)}}^{\text{HT}} (A_i,\bold{X}_i,\Theta_i;\boldsymbol{\eta},\beta^{(0)})\triangleq \sum_{i=1}^n \bigg(\frac{1-A_i}{1-\pi(\bold{X}_i)}\sin{\Theta}_i - \beta^{(0)} \bigg)=0.
\end{equation}
Combining (\ref{logistic EE}), (\ref{alpha 1 EE}) - (\ref{beta 0 EE}) leads to 
\begin{equation}\label{HT estimator EE}
    \sum_{i=1}^n \psi_{\boldsymbol{\vartheta}}^{\text{HT}}(A_i,\bold{X}_i,\Theta_i;\boldsymbol{\vartheta}) \triangleq\begin{bmatrix}
            &\sum\limits_{i=1}^n \psi_{\boldsymbol{\eta}}(A_i,\bold{X}_i;\boldsymbol{\eta}) \\
            &\sum\limits_{i=1}^n \psi_{\alpha^{(1)}}^{\text{HT}}(A_i,\bold{X}_i,\Theta_i;\boldsymbol{\eta},\alpha^{(1)}) \\
            &\sum\limits_{i=1}^n \psi_{\beta^{(1)}}^{\text{HT}}(A_i,\bold{X}_i,\Theta_i;\boldsymbol{\eta},\beta^{(1)}) \\
            &\sum\limits_{i=1}^n \psi_{\alpha^{(0)}}^{\text{HT}} (A_i,\bold{X}_i,\Theta_i;\boldsymbol{\eta},\alpha^{(0)})\\
            &\sum\limits_{i=1}^n \psi_{\beta^{(0)}}^{\text{HT}}(A_i,\bold{X}_i,\Theta_i;\boldsymbol{\eta},\beta^{(0)})
    \end{bmatrix} = \bold{0}_{5+p}.
\end{equation}
Therefore, the HT-type estimator $\widehat{\boldsymbol{\vartheta}}$ is the solution to (\ref{HT estimator EE}). By Stefanski and Boos (2002), as $n\to\infty$, the asymptotic distribution of this estimator is given by
\begin{equation}
    n^{1/2}(\widehat{\boldsymbol{\vartheta}}-\boldsymbol{\vartheta}) \overset{d}{\longrightarrow} \text{N}(\bold{0}_{5+p}, \gamma_{\text{HT}}(\boldsymbol{\vartheta})), 
\end{equation}
where $\gamma_\text{HT}(\boldsymbol{\vartheta}) = A_\text{HT}(\boldsymbol{\vartheta})^{-1} B_\text{HT}(\boldsymbol{\vartheta})^{-1} \{A_\text{HT}(\boldsymbol{\vartheta})^{-1} \}^{\top}$ with matrices 
\begin{equation}\label{A_1 matrix}
   \begin{aligned}
        A_\text{HT}(\boldsymbol{\vartheta}) &= -E\bigg\{\frac{\partial}{\partial\boldsymbol{\vartheta}^\top}\psi_{\boldsymbol{\vartheta}}^{\text{HT}}(A,\bold{X},\Theta;\boldsymbol{\vartheta})\bigg\}\\
        &=\begin{bmatrix}
            -E \bigg(\dfrac{\partial}{\partial\boldsymbol{\eta}^\top}\psi_{\boldsymbol{\eta}}\bigg)&\bold{0}_{1+p} &\bold{0}_{1+p}&\bold{0}_{1+p}&\bold{0}_{1+p} \\[0.25cm]
            -E \bigg(\dfrac{\partial}{\partial\boldsymbol{\eta}^\top} \psi_{\alpha^{(1)}}^{\text{HT}}\bigg) &-E\bigg(\dfrac{\partial}{\partial\alpha^{(1)}}\psi_{\alpha^{(1)}}^{\text{HT}}\bigg) &0&0&0\\[0.25cm]
            -E \bigg(\dfrac{\partial}{\partial\boldsymbol{\eta}^\top} \psi_{\beta^{(1)}}^{\text{HT}}\bigg) &0 &-E\bigg(\dfrac{\partial}{\partial\beta^{(1)}}\psi_{\beta^{(1)}}^{\text{HT}}\bigg) &0&0\\[0.25cm]
            -E \bigg(\dfrac{\partial}{\partial\boldsymbol{\eta}^\top} \psi_{\alpha^{(0)}}^{\text{HT}}\bigg) &0 &0&-E\bigg(\dfrac{\partial}{\partial\alpha^{(0)}}\psi_{\alpha^{(0)}}^{\text{HT}}\bigg) &0\\[0.25cm]
            -E \bigg(\dfrac{\partial}{\partial\boldsymbol{\eta}^\top}\psi_{\beta^{(1)}}^{\text{HT}}\bigg) &0 &0&0&E\bigg(\dfrac{\partial}{\partial\beta^{(0)}}\psi_{\beta^{(0)}}^{\text{HT}}\bigg) 
        \end{bmatrix}\\
        &=\left[\begin{array}{c|cccc}
-E \bigg(\dfrac{\partial}{\partial\boldsymbol{\eta}^\top}\psi_{\boldsymbol{\eta}}\bigg)&\bold{0}_{1+p} &\bold{0}_{1+p}&\bold{0}_{1+p}&\bold{0}_{1+p} \\[0.25cm]
            \hline
            -E \bigg(\dfrac{\partial}{\partial\boldsymbol{\eta}^\top} \psi_{\alpha^{(1)}}^{\text{HT}}\bigg) &1 &0&0&0\\[0.25cm]
            -E \bigg(\dfrac{\partial}{\partial\boldsymbol{\eta}^\top} \psi_{\beta^{(1)}}^{\text{HT}}\bigg) &0 &1 &0&0\\[0.25cm]
            -E \bigg(\dfrac{\partial}{\partial\boldsymbol{\eta}^\top} \psi_{\alpha^{(0)}}^{\text{HT}}\bigg) &0 &0&1&0\\[0.25cm]
            -E \bigg(\dfrac{\partial}{\partial\boldsymbol{\eta}^\top}\psi_{\beta^{(1)}}^{\text{HT}}\bigg) &0 &0&0&1
        \end{array}
        \right]\\
        &\triangleq \begin{bmatrix}
            \bold{a}_{11} & \bold{0}_{(1+p)\times4}\\
            \bold{a}_{21} & \bold{I}_{4}
        \end{bmatrix},
   \end{aligned}
\end{equation}
\begin{equation}\label{A_1 matrix inverse}
    \{A_\text{HT}(\boldsymbol{\vartheta})\}^{-1} = \begin{bmatrix}
        \bold{a}_{11}^{-1}& \bold{0}_{(1+p)\times4}\\
        -\bold{a}_{11}^{-1}\bold{a}_{21}& \bold{I}_{4}
    \end{bmatrix},
\end{equation}
\begin{equation}\label{B_1 matrix}
  \begin{aligned}
        B_\text{HT}(\boldsymbol{\vartheta}) &= E\bigg\{\psi_{\boldsymbol{\vartheta}}^{\text{HT}}(A,\bold{X},\Theta;\boldsymbol{\vartheta})\psi_{\boldsymbol{\vartheta}}^{\text{HT}}(A,\bold{X},\Theta;\boldsymbol{\vartheta})^\top\bigg\}\\
        &=E\left[\begin{array}{c|cccc}
            \psi_{\boldsymbol{\eta}}\psi_{\boldsymbol{\eta}}^\top &  \psi_{\boldsymbol{\eta}}\psi^{\text{HT}}_{\alpha^{(1)}}&  \psi_{\boldsymbol{\eta}}\psi^{\text{HT}}_{\beta^{(1)}}&  \psi_{\boldsymbol{\eta}}\psi^{\text{HT}}_{\alpha^{(0)}}&  \psi_{\boldsymbol{\eta}}\psi^{\text{HT}}_{\beta^{(0)}}\\
            \hline
            \psi^{\text{HT}}_{\alpha^{(1)}}\psi_{\boldsymbol{\eta}}^\top &\psi^{\text{HT}}_{\alpha^{(1)}}\psi^{\text{HT}}_{\alpha^{(1)}}&\psi^{\text{HT}}_{\alpha^{(1)}}\psi^{\text{HT}}_{\beta^{(1)}}&\psi^{\text{HT}}_{\alpha^{(1)}}\psi^{\text{HT}}_{\alpha^{(0)}}&\psi^{\text{HT}}_{\alpha^{(1)}}\psi^{\text{HT}}_{\beta^{(0)}}\\
            \psi^{\text{HT}}_{\beta^{(1)}}\psi_{\boldsymbol{\eta}}^\top &\psi^{\text{HT}}_{\beta^{(1)}}\psi^{\text{HT}}_{\alpha^{(1)}}&\psi^{\text{HT}}_{\beta^{(1)}}\psi^{\text{HT}}_{\beta^{(1)}}&\psi^{\text{HT}}_{\beta^{(1)}}\psi^{\text{HT}}_{\alpha^{(0)}}&\psi^{\text{HT}}_{\beta^{(1)}}\psi^{\text{HT}}_{\beta^{(0)}}\\
            \psi^{\text{HT}}_{\alpha^{(0)}}\psi_{\boldsymbol{\eta}}^\top &\psi^{\text{HT}}_{\alpha^{(0)}}\psi^{\text{HT}}_{\alpha^{(1)}}&\psi^{\text{HT}}_{\alpha^{(0)}}\psi^{\text{HT}}_{\beta^{(1)}}&\psi^{\text{HT}}_{\alpha^{(0)}}\psi^{\text{HT}}_{\alpha^{(0)}}&\psi^{\text{HT}}_{\alpha^{(0)}}\psi^{\text{HT}}_{\beta^{(0)}}\\
            \psi^{\text{HT}}_{\beta^{(1)}}\psi_{\boldsymbol{\eta}}^\top &\psi^{\text{HT}}_{\beta^{(0)}}\psi^{\text{HT}}_{\alpha^{(1)}}&\psi^{\text{HT}}_{\beta^{(0)}}\psi^{\text{HT}}_{\beta^{(1)}}&\psi^{\text{HT}}_{\beta^{(0)}}\psi^{\text{HT}}_{\alpha^{(0)}}&\psi^{\text{HT}}_{\beta^{(0)}}\psi^{\text{HT}}_{\beta^{(0)}}\\
        \end{array}\right]\\
        &\triangleq\begin{bmatrix}
            \bold{b}_{11} & \bold{b}_{21}^\top\\
             \bold{b}_{21} &  \bold{b}_{22}
        \end{bmatrix}
  \end{aligned}
\end{equation}
and $\bold{I}_k$ being the $k\times k$ identity matrix for a positive integer $k$. A detailed investigation of the components of the submatrices is provided in Appendix E of Wu and Chen (2025). Here, we present some core results in it to simplify the variance-covariance matrix $\gamma_{\text{HT}}(\boldsymbol{\vartheta})$. Firstly, owing to Lemma 7.3.11 of Casella and Berger (2002), we have that $\bold{a}_{11} = \bold{b}_{11}=E[\pi(\bold{X;\boldsymbol{\eta}})\{1-\pi(\bold{X;\boldsymbol{\eta}})\}\bold{X}^\top\bold{X}]$.

Secondly, we explore the relationship between $\bold{a}_{21}$ and $\bold{b}_{21}$. One one hand, we have that
$$
\begin{aligned}
    \bold{a}_{21} &\triangleq -\begin{bmatrix}
    E\bigg(\dfrac{\partial}{\partial\boldsymbol{\eta}^\top}\psi^\text{HT}_{\alpha^{(1)}}\bigg) & E\bigg(\dfrac{\partial}{\partial\boldsymbol{\eta}^\top}\psi^\text{HT}_{\beta^{(1)}}\bigg) & E\bigg(\dfrac{\partial}{\partial\boldsymbol{\eta}^\top}\psi^\text{HT}_{\alpha^{(0)}}\bigg)& E\bigg(\dfrac{\partial}{\partial\boldsymbol{\eta}^\top}\psi^\text{HT}_{\beta^{(0)}}\bigg) 
\end{bmatrix}^\top\\
&= \begin{bmatrix}
        &E\bigg(\dfrac{1-\pi(\bold{X})}{\pi(\bold{X})}A\cos{\Theta}\bold{X}\bigg)\\[0.25cm]
        & E\bigg(\dfrac{1-\pi(\bold{X})}{\pi(\bold{X})}A\sin{\Theta}\bold{X}\bigg)\\[0.25cm]
        & E\bigg(\dfrac{\pi(\bold{X})}{1-\pi(\bold{X})}(1-A)\cos{\Theta}\bold{X}\bigg)\\[0.25cm]
        & E\bigg(\dfrac{\pi(\bold{X})}{1-\pi(\bold{X})}(1-A)\sin{\Theta}\bold{X}\bigg)
    \end{bmatrix}\\
&=\begin{bmatrix}
    &E\bigg[\{1-\pi(\bold{X})\}\cos{\Theta^{(1)}}\bold{X}^\top\}\bigg] \\[0.25cm]
    &E\bigg[\{1-\pi(\bold{X})\}\sin{\Theta^{(1)}}\bold{X}^\top\}\bigg] \\[0.25cm]
    &E\bigg\{\pi(\bold{X})\cos{\Theta^{(0)}}\bold{X}^\top\bigg\} \\[0.25cm]
    &E\bigg\{\pi(\bold{X})\sin{\Theta^{(0)}}\bold{X}^\top\bigg\} \\[0.25cm]
\end{bmatrix},
\end{aligned}
$$
where the second step is established by Lemma \ref{IPW lemma}. On the other hand, for the submartix $\bold{b}_{21} \triangleq E\begin{bmatrix}
    \psi^{\text{HT}}_{\alpha^{(1)}}\psi^\top_{\boldsymbol{\eta}} & \psi^{\text{HT}}_{\beta^{(1)}}\psi^\top_{\boldsymbol{\eta}} &
    \psi^{\text{HT}}_{\alpha^{(0)}}\psi^\top_{\boldsymbol{\eta}} & \psi^{\text{HT}}_{\beta^{(0)}}\psi^\top_{\boldsymbol{\eta}}
\end{bmatrix}^\top$, we examine that
\begin{equation}\label{first component of b21}
    \begin{aligned}
    E(\psi^{\text{HT}}_{\alpha^{(1)}}\psi^\top_{\boldsymbol{\eta}}) &= E\Bigg[\bigg(\frac{A}{\pi(\bold{X};\boldsymbol{\eta})}\cos{\Theta} - \alpha^{(1)}\bigg) \bold{X}^\top\{A-\pi(\bold{X};\boldsymbol{\eta})\} \Bigg] \\
    &=E \bigg(\frac{A\bold{X}^\top\cos{\Theta}}{\pi(\bold{X;\boldsymbol{\eta}})}\bigg) - E  (A\bold{X}^\top\cos{\Theta}) - \alpha^{(1)}E(A\bold{X}^\top)+\alpha^{(1)}E\{\bold{X}^\top\pi(\bold{X};\boldsymbol{\eta})\}\\
    &=E\Bigg[\frac{A\bold{X}^\top\cos{\Theta}}{\pi(\bold{X;\boldsymbol{\eta}})}\{1-\pi(\bold{X;\boldsymbol{\eta}})\}  \Bigg]-\alpha^{(1)}E\{\bold{X}^\top\pi(\bold{X};\boldsymbol{\eta})\}+\alpha^{(1)}E\{\bold{X}^\top\pi(\bold{X};\boldsymbol{\eta})\}\\
    &=E\bigg[\{1-\pi(\bold{X})\}\cos{\Theta^{(1)}}\bold{X}^\top\}\bigg],
\end{aligned}
\end{equation}
where we utilizes the tower property of expectation on the third term of the second line and Lemma \ref{IPW lemma} on the first term of the third line. The remaining steps only involves basic calculations. With the similar derivation demonstrated in (\ref{first component of b21}), we can show that other components in $\bold{a}_{21}$ and $ \bold{b}_{21}$ are equal and hence $\bold{a}_{21} = \bold{b}_{21}$. 

Finally, we thoroughly investigate the components in $\bold{b}_{22}$. Consider $a,t\in\{0,1\}$. For $a=1$ and $t=0$, we have that
\begin{equation}\label{psi alpha 1 psi alpha 0}
    \begin{aligned}
    E(\psi^{\text{HT}}_{\alpha^{(1)}}\psi^{\text{HT}}_{\alpha^{(0)}}) &=  E\Bigg\{\bigg(\frac{A}{\pi(\bold{X})}\cos{\Theta}-\alpha^{(1)}\bigg)\bigg(\frac{1-A}{1-\pi(\bold{X})}\cos{\Theta}-\alpha^{(0)}\bigg)\Bigg\}\\
    &=E\Bigg\{\frac{A(1-A)}{\pi(\bold{X})\{1-\pi(\bold{X})\}}\cos^2{\Theta}\Bigg\} - \alpha^{(0)}E\bigg(\frac{A}{\pi(\bold{X})}\cos{\Theta}\bigg)\\
    &\;\;\;\;- \alpha^{(1)}E\bigg(\frac{1-A}{1-\pi(\bold{X})}\cos{\Theta}\bigg)+\alpha^{(1)}\alpha^{(0)} \\
    &= 0-\alpha^{(0)}\alpha^{(1)}-\alpha^{(1)}\alpha^{(0)}+\alpha^{(1)}\alpha^{(0)}\\
    &=-\alpha^{(1)}\alpha^{(0)},
\end{aligned}
\end{equation}
where the third equality is a result of $A(1-A)=0$ and Lemma \ref{IPW lemma} with $g(\Theta) = \cos{\Theta}$ and other equalities only involve basic calculation of expectations. With similar use of  $A(1-A)=0$ and Lemma \ref{IPW lemma}, one can generalize equation (\ref{psi alpha 1 psi alpha 0}) to the case $a\not=t$, which gives
\begin{align}
        E(\psi^{\text{HT}}_{\alpha^{(a)}}\psi^{\text{HT}}_{\alpha^{(t)}})  &=- \alpha^{(a)}\alpha^{(t)},\label{HT EE alpha a alpha t}\\
        E(\psi^{\text{HT}}_{\alpha^{(a)}}\psi^{\text{HT}}_{\beta^{(t)}})&=-\alpha^{(a)}\beta^{(t)}\label{HT EE alpha a beta t}
\end{align}
and
\begin{equation}\label{HT EE beta a beta t}
    E(\psi^{\text{HT}}_{\beta^{(a)}}\psi^{\text{HT}}_{\beta^{(t)}}) =-\beta^{(a)}\beta^{(t)}.
\end{equation}

Moreover, we consider the case where $a=t=1$.
\begin{equation}\label{derivation HT EE alpha1 alpha1}
    \begin{aligned}        E(\psi^{\text{HT}}_{\alpha^{(1)}}\psi^{\text{HT}}_{\alpha^{(1)}}) &= E\Bigg\{\bigg(\frac{A}{\pi(\bold{X})}\cos{\Theta}-\alpha^{(1)}\bigg)^2\Bigg\}\\
&=E\bigg(\frac{A}{\pi(\bold{X})^2}\cos^2{\Theta}\bigg) -2E\Bigg\{\bigg(\frac{A}{\pi(\bold{X})}\cos{\Theta}\bigg)\Bigg\}\alpha^{(1)}+\{\alpha^{(1)}\}^2\\
&=E\bigg(\frac{\cos^{2}{\Theta^{(1)}}}{\pi(\bold{X})}\bigg)-2\{\alpha^{(1)}\}^2+\{\alpha^{(1)}\}^2\\
&=E\bigg(\frac{\cos^{2}{\Theta^{(1)}}}{\pi(\bold{X})}\bigg)-\{\alpha^{(1)}\}^2.
\end{aligned}
\end{equation}
Following the similar procedure in (\ref{derivation HT EE alpha1 alpha1}), the results of other cases where $a=t$ are exhaustively given by
\begin{align}
    E(\psi^{\text{HT}}_{\alpha^{(0)}}\psi^{\text{HT}}_{\alpha^{(0)}}) &=E\bigg(\frac{\sin^{2}{\Theta^{(1)}}}{1-\pi(\bold{X})}\bigg)-\{\alpha^{(0)}\}^2,\label{HT EE alpha0 alpha0}\\
     E(\psi^{\text{HT}}_{\alpha^{(1)}}\psi^{\text{HT}}_{\beta^{(1)}}) &=E\bigg(\frac{\cos{\Theta^{(1)}}\sin{\Theta^{(1)}}}{\pi(\bold{X})}\bigg)-\alpha^{(1)}\beta^{(1)},\\
     E(\psi^{\text{HT}}_{\alpha^{(0)}}\psi^{\text{HT}}_{\beta^{(0)}})
     &=E\bigg(\frac{\cos{\Theta^{(0)}}\sin{\Theta^{(0)}}}{1-\pi(\bold{X})}\bigg)-\alpha^{(0)}\beta^{(0)},\\
     E(\psi^{\text{HT}}_{\beta^{(1)}}\psi^{\text{HT}}_{\beta^{(1)}}) &=E\bigg(\frac{\sin^{2}{\Theta^{(1)}}}{\pi(\bold{X})}\bigg)-\{\beta^{(1)}\}^2\\
     \intertext{and}
E(\psi^{\text{HT}}_{\beta^{(0)}}\psi^{\text{HT}}_{\beta^{(0)}})&=E\bigg(\frac{\sin^{2}{\Theta^{(0)}}}{1-\pi(\bold{X})}\bigg)-\{\beta^{(0)}\}^2\label{HT EE beta0 beta0}.
\end{align}
 With all components determined in (\ref{HT EE alpha a alpha t})-(\ref{HT EE beta0 beta0}), it follows that
\begin{equation}
   \begin{aligned}
        \bold{b}_{22} &=\begin{bmatrix}
        E\bigg(\dfrac{\cos^{2}{\Theta^{(1)}}}{\pi(\bold{X})}\bigg) &E\bigg(\dfrac{\cos{\Theta^{(1)}}\sin{\Theta^{(1)}}}{\pi(\bold{X})}\bigg) &0 &0\\
         E\bigg(\dfrac{\cos{\Theta^{(1)}}\sin{\Theta^{(1)}}}{\pi(\bold{X})}\bigg) &E\bigg(\dfrac{\sin^{2}{\Theta^{(1)}}}{\pi(\bold{X})}\bigg) &0&0\\
         0 &0 &E\bigg(\dfrac{\sin^{2}{\Theta^{(1)}}}{1-\pi(\bold{X})}\bigg)&E\bigg(\dfrac{\cos{\Theta^{(0)}}\sin{\Theta^{(0)}}}{1-\pi(\bold{X})}\bigg)\\
         0&0&E\bigg(\dfrac{\cos{\Theta^{(0)}}\sin{\Theta^{(0)}}}{1-\pi(\bold{X})}\bigg)&E\bigg(\dfrac{\sin^{2}{\Theta^{(0)}}}{1-\pi(\bold{X})}\bigg) 
    \end{bmatrix}\\
    \;\;\;\;&-\begin{bmatrix}
        \alpha^{(1)}\alpha^{(1)} & \alpha^{(1)}\beta^{(1)} &\alpha^{(1)}\alpha^{(0)} &\alpha^{(1)}\beta^{(0)}\\
        \alpha^{(1)}\beta^{(1)} &\beta^{(1)}\beta^{(1)}& \beta^{(1)}\alpha^{(0)} & \beta^{(0)}\beta^{(1)}\\
        \alpha^{(1)}\alpha^{(0)} &\alpha^{(1)}\beta^{(0)}&\alpha^{(0)}\alpha^{(0)} &\alpha^{(0)}\beta^{(0)}\\
        \beta^{(1)}\alpha^{(0)} & \beta^{(0)}\beta^{(1)}&\alpha^{(0)}\beta^{(0)} &\beta^{(0)} \beta^{(0)} 
    \end{bmatrix}\\
    &=\begin{bmatrix}
        \mathbb{B}^{(1)} &\bold{0}_{2\times2}\\
        \bold{0}_{2\times2} &\mathbb{B}^{(0)}
    \end{bmatrix} - \boldsymbol{\omega}\boldsymbol{\omega}^{\top}
   \end{aligned}
\end{equation}
where
\begin{align}
    \mathbb{B}^{(1)} &\triangleq \begin{bmatrix}
    E\bigg(\dfrac{\cos^{2}{\Theta^{(1)}}}{\pi(\bold{X})}\bigg) & E\bigg(\dfrac{\cos{\Theta^{(1)}}\sin{\Theta^{(1)}}}{\pi(\bold{X})}\bigg)\\
    E\bigg(\dfrac{\cos{\Theta^{(1)}}\sin{\Theta^{(1)}}}{\pi(\bold{X})}\bigg)& E\bigg(\dfrac{\sin^{2}{\Theta^{(1)}}}{\pi(\bold{X})}\bigg)
\end{bmatrix},
 \intertext{and}
 \mathbb{B}^{(0)} &\triangleq \begin{bmatrix}
     E\bigg(\dfrac{\sin^{2}{\Theta^{(1)}}}{1-\pi(\bold{X})}\bigg) & E\bigg(\dfrac{\cos{\Theta^{(0)}}\sin{\Theta^{(0)}}}{1-\pi(\bold{X})}\bigg) \\
     E\bigg(\dfrac{\cos{\Theta^{(0)}}\sin{\Theta^{(0)}}}{1-\pi(\bold{X})}\bigg) & E\bigg(\dfrac{\sin^{2}{\Theta^{(0)}}}{1-\pi(\bold{X})}\bigg)
 \end{bmatrix}.
\end{align}
With (\ref{A_1 matrix inverse}) and (\ref{B_1 matrix}), we express the variance-covariance matrix as 
\begin{equation}\label{var-cov HT}
    \begin{aligned}
        \gamma_{\text{HT}}(\boldsymbol{\vartheta}) &= \begin{bmatrix}
        \bold{a}_{11}^{-1}\bold{b}_{11}\bold{a}_{11}^{-1} & - \bold{a}_{11}^{-1}(\bold{a}_{21}^\top-\bold{b}_{21}^\top)\\
        (-\bold{a}_{21}+\bold{b}_{21})\bold{a}_{11}^{-1} &(\bold{a}_{21}-\bold{b}_{21})\bold{a}_{11}^{-1}\bold{a}_{21}^\top - \bold{a}_{21}\bold{a}_{11}^{-1}\bold{b}_{21}^\top+\bold{b}_{22} 
    \end{bmatrix}\\
        &= \begin{bmatrix}
            \bold{a}_{11}^{-1} & \bold{0}_{(1+p)\times4}\\
            \bold{0}_{4\times(1+p)} &-\bold{b}_{21}\bold{a}_{11}^{-1}\bold{b}_{21}^\top+\bold{b}_{22} 
        \end{bmatrix}.
    \end{aligned}
\end{equation}
By the property of multivariate normal distribution, we have that
$$
n^{1/2}(\widehat{\boldsymbol{\eta}}-\boldsymbol{\eta})\overset{d}{\longrightarrow}\text{N}(\bold{0}_{1+p},\mathcal{A}^{-1}),
$$
and
$$
n^{1/2}(\widehat{\boldsymbol{\omega}} - \boldsymbol{\omega})\overset{d}{\longrightarrow}\text{N}(\bold{0}_4,-\mathcal{B}\mathcal{A}\mathcal{B}^\top + \mathcal{C})
$$
where $\mathcal{A}\triangleq\bold{a}^{-1}_{11}$, $\mathcal{B}\triangleq\bold{b}_{21}$ and $\mathcal{C}\triangleq\bold{b}_{22}$. This gives results (i) and (ii).
\\
\textit{\underline{Part 2: H\'ajek-type estimators}}
\\
The procedures of deriving asymptotics for H\'ajek-type estimators are similar with Part 1 of the proof. The only difference is that the estimating equations are different. For H\'ajek-type estimators (\ref{Hajek alpha}) and (\ref{Hajek beta}) with $a\in\{0,1\}$, they satisfies the following equations 
\begin{align}
    \sum_{i=1}^n\psi_{\alpha^{(1)}}^\text{H\'ajek} (A_i,\bold{X}_i,\Theta_i;\boldsymbol{\eta},\alpha^{(1)}) &\triangleq \sum_{i=1}^n \frac{A_i}{\pi(\bold{X}_i)}(\cos{\Theta}_i-\alpha^{(1)})=0, \label{Hajek EE alpha 1}\\
     \sum_{i=1}^n\psi_{\beta^{(1)}}^\text{H\'ajek} (A_i,\bold{X}_i,\Theta_i;\boldsymbol{\eta},\beta^{(1)}) &\triangleq \sum_{i=1}^n\frac{A_i}{\pi(\bold{X}_i)}(\sin{\Theta}_i-\beta^{(1)})=0,\\
     \sum_{i=1}^n\psi_{\alpha^{(0)}}^\text{H\'ajek} (A_i,\bold{X}_i,\Theta_i;\boldsymbol{\eta},\alpha^{(0)}) &\triangleq   \sum_{i=1}^n \frac{1-A_i}{1-\pi(\bold{X}_i)}(\cos{\Theta}_i-\alpha^{(0)})=0\\
     \intertext{and}
     \sum_{i=1}^n\psi_{\beta^{(0)}}^\text{H\'ajek} (A_i,\bold{X}_i,\Theta_i;\boldsymbol{\eta},\beta^{(0)}) &\triangleq \sum_{i=1}^n \frac{1-A_i}{1-\pi(\bold{X}_i)}(\sin{\Theta}_i-\beta^{(0)})=0\label{Hajek EE beta 0}.
\end{align}
Since the treatment model is handled by logistic regression,  the equation (\ref{logistic EE}) is directly adopted in forming the estimating equations. Combining (\ref{logistic EE}) and (\ref{Hajek EE alpha 1}) - (\ref{Hajek EE beta 0}) gives 
\begin{equation}\label{Hajek EE}
    \sum_{i=1}^n \psi_{\boldsymbol{\vartheta}}^{\text{H\'ajek}} (A_i,\bold{X}_i,\Theta_i;\boldsymbol{\vartheta}) = \begin{bmatrix}
        \sum\limits_{i=1}^n \psi_{\boldsymbol{\eta}} (A_i,\bold{X}_i;\boldsymbol{\eta}) \\
         \sum\limits_{i=1}^n\psi_{\alpha^{(1)}}^\text{H\'ajek} (A_i,\bold{X}_i,\Theta_i;\boldsymbol{\eta},\alpha^{(1)}) \\
         \sum\limits_{i=1}^n\psi_{\beta^{(1)}}^\text{H\'ajek} (A_i,\bold{X}_i,\Theta_i;\boldsymbol{\eta},\beta^{(1)})\\
          \sum\limits_{i=1}^n\psi_{\alpha^{(0)}}^\text{H\'ajek} (A_i,\bold{X}_i,\Theta_i;\boldsymbol{\eta},\alpha^{(0)}) \\
          \sum\limits_{i=1}^n\psi_{\beta^{(0)}}^\text{H\'ajek} (A_i,\bold{X}_i,\Theta_i;\boldsymbol{\eta},\beta^{(0)})
    \end{bmatrix} = \bold{0}_{5+p},
\end{equation}
to which $\widetilde{\boldsymbol{\vartheta}}$ is a solution. According to Stefanski and Boos (2002), we have the asymptotic distribution as 
\begin{equation}
    n^{1/2}(\widetilde{\boldsymbol{\vartheta}} - \boldsymbol{\vartheta})\overset{d}{\longrightarrow} \text{N}(\bold{0}_{5+p},\gamma_{\text{H\'ajek}}(\boldsymbol{\vartheta})),
\end{equation}
where $\gamma_{\text{H\'ajek}}(\boldsymbol{\vartheta})= A_{\text{H\'ajek}}(\boldsymbol{\vartheta})^{-1}B_{\text{H\'ajek}}(\boldsymbol{\vartheta})\{A_{\text{H\'ajek}}(\boldsymbol{\vartheta})^{-1}\}^\top$ with matrices given by
\begin{equation}
    \begin{aligned}
        A_{\text{H\'ajek}}(\boldsymbol{\vartheta}) &=
        -E\bigg\{\frac{\partial}{\partial\boldsymbol{\vartheta}^\top}\psi_{\boldsymbol{\vartheta}}^{\text{H\'ajek}} (A,\bold{X},\Theta;\boldsymbol{\vartheta})\bigg\}\\
        &=\begin{adjustbox}{max width=\linewidth}
            $\begin{bmatrix}
            -E \bigg(\dfrac{\partial}{\partial\boldsymbol{\eta}^\top}\psi_{\boldsymbol{\eta}}\bigg)&\bold{0}_{1+p} &\bold{0}_{1+p}&\bold{0}_{1+p}&\bold{0}_{1+p} \\[0.25cm]
            -E \bigg(\dfrac{\partial}{\partial\boldsymbol{\eta}^\top} \psi_{\alpha^{(1)}}^{\text{H\'ajek}}\bigg) &-E\bigg(\dfrac{\partial}{\partial\alpha^{(1)}}\psi_{\alpha^{(1)}}^{\text{H\'ajek}}\bigg) &0&0&0\\[0.25cm]
            -E \bigg(\dfrac{\partial}{\partial\boldsymbol{\eta}^\top} \psi_{\beta^{(1)}}^{\text{H\'ajek}}\bigg) &0 &-E\bigg(\dfrac{\partial}{\partial\beta^{(1)}}\psi_{\beta^{(1)}}^{\text{H\'ajek}}\bigg) &0&0\\[0.25cm]
            -E \bigg(\dfrac{\partial}{\partial\boldsymbol{\eta}^\top} \psi_{\alpha^{(0)}}^{\text{H\'ajek}}\bigg) &0 &0&-E\bigg(\dfrac{\partial}{\partial\alpha^{(0)}}\psi_{\alpha^{(0)}}^{\text{H\'ajek}}\bigg) &0\\[0.25cm]
            -E \bigg(\dfrac{\partial}{\partial\boldsymbol{\eta}^\top}\psi_{\beta^{(1)}}^{\text{H\'ajek}}\bigg) &0 &0&0&E\bigg(\dfrac{\partial}{\partial\beta^{(0)}}\psi_{\beta^{(0)}}^{\text{H\'ajek}}\bigg) 
        \end{bmatrix}$
        \end{adjustbox}\\
        &=\left[\begin{array}{c|cccc}
           -E \bigg(\dfrac{\partial}{\partial\boldsymbol{\eta}^\top}\psi_{\boldsymbol{\eta}}\bigg)&\bold{0}_{1+p} &\bold{0}_{1+p}&\bold{0}_{1+p}&\bold{0}_{1+p} \\[0.25cm]\hline
            -E \bigg(\dfrac{\partial}{\partial\boldsymbol{\eta}^\top} \psi_{\alpha^{(1)}}^{\text{H\'ajek}}\bigg) &1 &0&0&0\\[0.25cm]
            -E \bigg(\dfrac{\partial}{\partial\boldsymbol{\eta}^\top} \psi_{\beta^{(1)}}^{\text{H\'ajek}}\bigg) &0 &1 &0&0\\[0.25cm]
            -E \bigg(\dfrac{\partial}{\partial\boldsymbol{\eta}^\top} \psi_{\alpha^{(0)}}^{\text{H\'ajek}}\bigg) &0 &0&1&0\\[0.25cm]
            -E \bigg(\dfrac{\partial}{\partial\boldsymbol{\eta}^\top}\psi_{\beta^{(1)}}^{\text{H\'ajek}}\bigg) &0 &0&0&1 
        \end{array}\right]\\
        &=\begin{bmatrix}
            \bold{a}_{11} &\bold{0}_{4\times(1+p)}\\
            \bold{\mathfrak{a}}_{21}&\bold{I}_4
        \end{bmatrix},
    \end{aligned}
\end{equation}
\begin{equation}
    A_{\text{H\'ajek}}(\boldsymbol{\vartheta})^{-1} = \begin{bmatrix}
        \bold{a}_{11}^{-1} & \bold{0}_{4\times(1+p)}\\
        - \bold{a}_{11}^{-1}\bold{\mathfrak{a}}_{21} &\bold{I}_4
    \end{bmatrix}
\end{equation}
and
\begin{eqnarray}\begin{aligned}
    B_{\text{H\'ajek}}(\boldsymbol{\vartheta}) &= 
        E\bigg\{\psi_{\boldsymbol{\vartheta}}^{\text{H\'ajek}} (A,\bold{X},\Theta;\boldsymbol{\vartheta})\psi_{\boldsymbol{\vartheta}}^{\text{H\'ajek}} (A,\bold{X},\Theta;\boldsymbol{\vartheta})^\top\bigg\}\\
        &=E\left[\begin{array}{c|cccc}
            \psi_{\boldsymbol{\eta}}\psi_{\boldsymbol{\eta}}^\top &  \psi_{\boldsymbol{\eta}}\psi^{\text{H\'ajek}}_{\alpha^{(1)}}&  \psi_{\boldsymbol{\eta}}\psi^{\text{H\'ajek}}_{\beta^{(1)}}&  \psi_{\boldsymbol{\eta}}\psi^{\text{H\'ajek}}_{\alpha^{(0)}}&  \psi_{\boldsymbol{\eta}}\psi^{\text{H\'ajek}}_{\beta^{(0)}}\\
            \hline
            \psi^{\text{H\'ajek}}_{\alpha^{(1)}}\psi_{\boldsymbol{\eta}}^\top &\psi^{\text{H\'ajek}}_{\alpha^{(1)}}\psi^{\text{H\'ajek}}_{\alpha^{(1)}}&\psi^{\text{H\'ajek}}_{\alpha^{(1)}}\psi^{\text{H\'ajek}}_{\beta^{(1)}}&\psi^{\text{H\'ajek}}_{\alpha^{(1)}}\psi^{\text{H\'ajek}}_{\alpha^{(0)}}&\psi^{\text{H\'ajek}}_{\alpha^{(1)}}\psi^{\text{H\'ajek}}_{\beta^{(0)}}\\
            \psi^{\text{H\'ajek}}_{\beta^{(1)}}\psi_{\boldsymbol{\eta}}^\top &\psi^{\text{H\'ajek}}_{\beta^{(1)}}\psi^{\text{H\'ajek}}_{\alpha^{(1)}}&\psi^{\text{H\'ajek}}_{\beta^{(1)}}\psi^{\text{H\'ajek}}_{\beta^{(1)}}&\psi^{\text{H\'ajek}}_{\beta^{(1)}}\psi^{\text{H\'ajek}}_{\alpha^{(0)}}&\psi^{\text{H\'ajek}}_{\beta^{(1)}}\psi^{\text{H\'ajek}}_{\beta^{(0)}}\\
            \psi^{\text{H\'ajek}}_{\alpha^{(0)}}\psi_{\boldsymbol{\eta}}^\top &\psi^{\text{H\'ajek}}_{\alpha^{(0)}}\psi^{\text{H\'ajek}}_{\alpha^{(1)}}&\psi^{\text{H\'ajek}}_{\alpha^{(0)}}\psi^{\text{H\'ajek}}_{\beta^{(1)}}&\psi^{\text{H\'ajek}}_{\alpha^{(0)}}\psi^{\text{H\'ajek}}_{\alpha^{(0)}}&\psi^{\text{H\'ajek}}_{\alpha^{(0)}}\psi^{\text{H\'ajek}}_{\beta^{(0)}}\\
            \psi^{\text{H\'ajek}}_{\beta^{(1)}}\psi_{\boldsymbol{\eta}}^\top &\psi^{\text{H\'ajek}}_{\beta^{(0)}}\psi^{\text{H\'ajek}}_{\alpha^{(1)}}&\psi^{\text{H\'ajek}}_{\beta^{(0)}}\psi^{\text{H\'ajek}}_{\beta^{(1)}}&\psi^{\text{H\'ajek}}_{\beta^{(0)}}\psi^{\text{H\'ajek}}_{\alpha^{(0)}}&\psi^{\text{H\'ajek}}_{\beta^{(0)}}\psi^{\text{H\'ajek}}_{\beta^{(0)}}\\
        \end{array}\right]\\
        &=\begin{bmatrix}
            \bold{{b}}_{11} & \bold{\mathfrak{b}}_{21}^\top\\
             \bold{\mathfrak{b}}_{21} &  \bold{\mathfrak{b}}_{22}
        \end{bmatrix}.
    \end{aligned}
\end{eqnarray}
In the spirit of Wu and Chen (2025), we have that 
\begin{equation}\label{a11 = b11  Hajek}
   \begin{aligned}
        \bold{\mathfrak{a}_{21}} = \bold{\mathfrak{b}_{21}} &= \begin{bmatrix}
           E\bigg[ \{1-\pi(\bold{X})\}\cos{(\Theta^{(1)}-\alpha^{(1)})}\bold{X}\bigg]\\[0.25cm]
             E\bigg[\{1-\pi(\bold{X})\}\sin{(\Theta^{(1)}-  \beta^{(1)})}\bold{X}\bigg]\\[0.25cm]
              E\bigg[\pi(\bold{X})\cos{(\Theta^{(0)}-  \alpha^{(0)})}\bold{X}\bigg]\\[0.25cm]
              E\bigg[\pi(\bold{X})\sin{(\Theta^{(0)}-  \beta^{(0)})}\bold{X}\bigg]
        \end{bmatrix}.
   \end{aligned}
\end{equation}
In Part 1, the result $\bold{a}_{11} = \bold{b}_{11}$ is already given. Combining these results gives the simplification of the asymptotic variance-covariance matrix, which is expressed as
\begin{equation}
    \gamma_{\text{H\'ajek}}(\boldsymbol{\vartheta}) = \begin{bmatrix}
    \bold{a}_{11}^{-1} & \bold{0}_{(1+p)\times4}\\
    \bold{0}_{4\times(1+p)} & -\bold{\mathfrak{b}}_{21}\bold{a}_{11}^{-1}\bold{\mathfrak{b}}_{21}^\top+\bold{\mathfrak{b}}_{22}
\end{bmatrix}.
\end{equation}
We take a closer look at $\bold{\mathfrak{b}}_{22}$, which is defined as 
\begin{equation}\label{b22 hajek}
    \begin{aligned}
        \bold{\mathfrak{b}}_{22} &\triangleq \begin{bmatrix}
        \psi^{\text{H\'ajek}}_{\alpha^{(1)}}\psi^{\text{H\'ajek}}_{\alpha^{(1)}}&\psi^{\text{H\'ajek}}_{\alpha^{(1)}}\psi^{\text{H\'ajek}}_{\beta^{(1)}}&\psi^{\text{H\'ajek}}_{\alpha^{(1)}}\psi^{\text{H\'ajek}}_{\alpha^{(0)}}&\psi^{\text{H\'ajek}}_{\alpha^{(1)}}\psi^{\text{H\'ajek}}_{\beta^{(0)}}\\
            \psi^{\text{H\'ajek}}_{\beta^{(1)}}\psi^{\text{H\'ajek}}_{\alpha^{(1)}}&\psi^{\text{H\'ajek}}_{\beta^{(1)}}\psi^{\text{H\'ajek}}_{\beta^{(1)}}&\psi^{\text{H\'ajek}}_{\beta^{(1)}}\psi^{\text{H\'ajek}}_{\alpha^{(0)}}&\psi^{\text{H\'ajek}}_{\beta^{(1)}}\psi^{\text{H\'ajek}}_{\beta^{(0)}}\\
            \psi^{\text{H\'ajek}}_{\alpha^{(0)}}\psi^{\text{H\'ajek}}_{\alpha^{(1)}}&\psi^{\text{H\'ajek}}_{\alpha^{(0)}}\psi^{\text{H\'ajek}}_{\beta^{(1)}}&\psi^{\text{H\'ajek}}_{\alpha^{(0)}}\psi^{\text{H\'ajek}}_{\alpha^{(0)}}&\psi^{\text{H\'ajek}}_{\alpha^{(0)}}\psi^{\text{H\'ajek}}_{\beta^{(0)}}\\
            \psi^{\text{H\'ajek}}_{\beta^{(0)}}\psi^{\text{H\'ajek}}_{\alpha^{(1)}}&\psi^{\text{H\'ajek}}_{\beta^{(0)}}\psi^{\text{H\'ajek}}_{\beta^{(1)}}&\psi^{\text{H\'ajek}}_{\beta^{(0)}}\psi^{\text{H\'ajek}}_{\alpha^{(0)}}&\psi^{\text{H\'ajek}}_{\beta^{(0)}}\psi^{\text{H\'ajek}}_{\beta^{(0)}}
    \end{bmatrix}\\
    &=\begin{bmatrix}
         \bold{\mathfrak{B}}^{(1)} &\bold{0}_{2\times2}\\
         \bold{0}_{2\times2} & \bold{\mathfrak{B}}^{(0)}
    \end{bmatrix},
    \end{aligned}
\end{equation}
where 
\begin{equation}\label{b22 b1 hajek}
   \bold{\mathfrak{B}}^{(1)} \triangleq\begin{bmatrix}
        E\bigg\{\dfrac{(\cos{\Theta^{(1)}}-\alpha^{(1)})^2}{\pi(\bold{X})}\bigg\}& E\bigg\{\dfrac{(\cos{\Theta^{(1)}}-\alpha^{(1)})(\sin{\Theta^{(1)}}-\beta^{(1)})}{\pi(\bold{X})}\bigg\} \\[0.25cm]
         E\bigg\{\dfrac{(\cos{\Theta^{(1)}}-\alpha^{(1)})(\sin{\Theta^{(1)}}-\beta^{(1)})}{\pi(\bold{X})}\bigg\} & E\bigg\{\dfrac{(\sin{\Theta^{(1)}}-\beta^{(1)})^2}{\pi(\bold{X})}\bigg\} 
   \end{bmatrix}
\end{equation}
and
\begin{equation}\label{b22 b2 hajek}
    \bold{\mathfrak{B}}^{(0)} \triangleq \begin{bmatrix}
         E\bigg\{\dfrac{(\cos{\Theta^{(0)}}-\alpha^{(0)})^2}{1-\pi(\bold{X})}\bigg\}& E\bigg\{\dfrac{(\cos{\Theta^{(0)}}-\alpha^{(0)})(\sin{\Theta^{(0)}}-\beta^{(0)})}{1-\pi(\bold{X})}\bigg\} \\[0.25cm]
         E\bigg\{\dfrac{(\cos{\Theta^{(0)}}-\alpha^{(0)})(\sin{\Theta^{(0)}}-\beta^{(0)})}{1-\pi(\bold{X})}\bigg\} & E\bigg\{\dfrac{(\sin{\Theta^{(0)}}-\beta^{(0)})^2}{1-\pi(\bold{X})}\bigg\} 
    \end{bmatrix}.
\end{equation}
Owing to the property of multivariate normal distribution, we obtain that 
$$
    n^{1/2}(\widetilde{\boldsymbol{\omega}} - \boldsymbol{\omega}) \overset{d}{\longrightarrow}\text{N}(\bold{0}_4,-\bold{\mathfrak{b}}_{21}\bold{a}_{11}^{-1}\bold{\mathfrak{b}}_{21}^\top+\bold{\mathfrak{b}}_{22})
$$
with the components of variance-covariance given in (\ref{a11 = b11  Hajek}), (\ref{b22 hajek})-(\ref{b22 b2 hajek}). Hence, the result (ii) is proved. \hfill$\square$
\section{Proof of Theorem \ref{parameters of interest asymptotics}}\label{parameters of interest proof}
Define
\begin{equation}\label{function for estimators}
    \boldsymbol{f}(\boldsymbol{\omega}) \triangleq \begin{bmatrix}
        \text{atan2}(\beta^{(1)},\alpha^{(1)}) - \text{atan2}(\beta^{(0)},\alpha^{(0)})  \\
        [\{\alpha^{(1)}\}^2 + \{\beta^{(1)}\}^2 ]^{1/2} - \ [\{\alpha^{(0)}\}^2 + \{\beta^{(0)}\}^2 ]^{1/2}
    \end{bmatrix}.
\end{equation} 
Note that (\ref{function for estimators}) gives $\widehat{\Delta} = \boldsymbol{f}(\widehat{\boldsymbol{\omega}}),\widetilde{\Delta} = \boldsymbol{f}(\widetilde{\boldsymbol{\omega}})$ and $\Delta = \boldsymbol{f}(\boldsymbol{\omega})$. The Jacobian matrix of (\ref{function for estimators}) is derived by
$$
   \begin{aligned}
        \bold{J} &\triangleq \begin{bmatrix}
        \dfrac{-\beta^{(1)}}{(\alpha^{(1)})^2+(\beta^{(1)})^2} & \dfrac{\alpha^{(1)}}{(\alpha^{(1)})^2+(\beta^{(1)})^2} & \dfrac{\beta^{(0)}}{(\alpha^{(0)})^2+(\beta^{(0)})^2} & \dfrac{-\alpha^{(0)}}{(\alpha^{(0)})^2+(\beta^{(0)})^2}\\[0.25cm]
          \dfrac{\alpha^{(1)}}{\{(\alpha^{(1)})^2+(\beta^{(1)})^2\}^{0.5}} & \dfrac{\beta^{(1)}}{\{(\alpha^{(1)})^2+(\beta^{(1)})^2\}^{0.5}} & \dfrac{-\alpha^{(0)}}{\{(\alpha^{(0)})^2+(\beta^{(0)})^2\}^{0.5}} & \dfrac{-\beta^{(0)}}{\{(\alpha^{(0)})^2+(\beta^{(0)})^2\}^{0.5}}\\
    \end{bmatrix}\\
    &=\left[\begin{array}{cc|cc}
     \dfrac{-\beta^{(1)}}{\{\rho^{(1)}\}^{2}}& \dfrac{\alpha^{(1)}}{\{\rho^{(1)}\}^2}& \dfrac{\beta^{(0)}}{\{\rho^{(0)}\}^2} & \dfrac{-\alpha^{(0)}}{\{\rho^{(0)}\}^2}\\[0.25cm]
          \dfrac{\alpha^{(1)}}{\rho^{(1)}} & \dfrac{\beta^{(1)}}{\rho^{(1)}}& \dfrac{-\alpha^{(0)}}{\rho^{(0)}} & \dfrac{-\beta^{(0)}}{\rho^{(0)}}
    \end{array}\right]\\
    &\triangleq\begin{bmatrix}
        \bold{D}^{(1)}\bold{\mathbb{J}}^{(1)} & \bold{D}^{(0)}\bold{\mathbb{J}}^{(0)} 
    \end{bmatrix},
   \end{aligned}
$$
where
$$
\begin{aligned}
    \mathbb{J}^{(1)} &\triangleq \begin{bmatrix}
        -\beta^{(1)} & \alpha^{(1)} \\
         \alpha^{(1)}&\beta^{(1)}
         \end{bmatrix},\\
          \mathbb{J}^{(0)} &\triangleq \begin{bmatrix}
        \beta^{(0)} & -\alpha^{(0)} \\
         -\alpha^{(0)}&-\beta^{(0)}
         \end{bmatrix}
         \end{aligned}
$$
and, for $a\in\{0,1\}$,
$$
    \bold{D}^{(a)} \triangleq \begin{bmatrix}
        \{\rho^{(a)}\}^{-2} &0\\
        0&\{\rho^{(a)}\}^{-1}
    \end{bmatrix}.
$$
Directly applying multivariate delta method on HT-type estimators yields
$$
    n^{1/2}(\widehat{\boldsymbol{\Delta}} - \boldsymbol{\Delta}) \overset{d}{\longrightarrow} \text{N}(\bold{0}_2,\boldsymbol{\Sigma}_{\text{HT}}),
$$
where
$$
    \begin{aligned}
            \boldsymbol{\Sigma}_{\text{HT}}&\triangleq\bold{J}(-\bold{b}_{21}\bold{a}_{11}^{-1}\bold{b}_{21}^\top+\bold{b}_{22})\bold{J}^\top \\
            &= -\bold{J}\bold{b}_{21}\bold{a}_{11}^{-1}\bold{b}_{21}^\top\bold{J}^\top + \bold{J}\bold{b}_{22}\bold{J}^\top\\
            &= -\mathcal{J}_{\text{HT}}\mathcal{A}^{-1}\mathcal{J}_{\text{HT}}^{\top}+ \bold{D}^{(1)}\mathbb{J}^{(1)}(\mathbb{B}^{(1)}-\boldsymbol{\omega}_1\boldsymbol{\omega}_1^\top)\mathbb{J}^{(1)}\bold{D}^{(1)}\\
            &\;\;\;\;+\bold{D}^{(0)}\mathbb{J}^{(0)}(\mathbb{B}^{(0)}-\boldsymbol{\omega}_0\boldsymbol{\omega}_0^\top)\mathbb{J}^{(0)}\bold{D}^{(0)}-2\bold{D}^{(1)}\mathbb{J}^{(1)}\boldsymbol{\omega}_1\boldsymbol{\omega}_0^\top\mathbb{J}^{(0)}\bold{D}^{(0)}
    \end{aligned}
$$
with $\mathcal{J}_{\text{HT}}\triangleq\bold{J}\bold{b}_{21},\boldsymbol{\omega}_1 \triangleq (\alpha^{(1)},\beta^{(1)})^\top$ and $\boldsymbol{\omega}_0 \triangleq(\alpha^{(0)},\beta^{(0)})^\top$. This leads to result (i). Again, applying multivariate delta method on H\'ajek-type estimators gives
$$
    n^{1/2}(\widetilde{\boldsymbol{\Delta}} - \boldsymbol{\Delta}) \overset{d}{\longrightarrow} \text{N}(\bold{0}_2,\boldsymbol{\Sigma}_{\text{H\'ajek}}),
$$
where
$$
 \begin{aligned}
       \boldsymbol{\Sigma}_{\text{H\'ajek}}&\triangleq \bold{J}(-\bold{\mathfrak{b}}_{21}\bold{a}_{11}^{-1}\bold{\mathfrak{b}}_{21}^\top+\bold{\mathfrak{b}}_{22})\bold{J}^\top\\ &= -\bold{J}\bold{\mathfrak{b}}_{21}\bold{a}_{11}^{-1}\bold{\mathfrak{b}}_{21}^\top\bold{J}^\top +\bold{J}\bold{\mathfrak{b}}_{22}\bold{J}^\top \\
        &=-\mathcal{J}_{\text{H\'ajek}}\mathcal{A}\mathcal{J}_{\text{H\'ajek}}^\top+\bold{D}^{(1)}\mathbb{J}^{(1)}\mathfrak{B}^{(1)}\mathbb{J}^{(1)}\bold{D}^{(1)}+\bold{D}^{(0)}\mathbb{J}^{(0)}\mathfrak{B}^{(0)}\mathbb{J}^{(0)}\bold{D}^{(0)}
    \end{aligned}
$$
with $\mathcal{J}_{\text{H\'ajek}}\triangleq \bold{J}\mathfrak{b}_{21}$. This brings about result (ii). \hfill$\square$
\section*{References}

\refmark{Casella, G. and Berger, R. L. (2002). \textit{Statistical Inference, Second Edition}. Duxbury Pacific Grove, CA.}

\refmark{Stefanski, L. A. and Boos, D. D. (2002). The calculus of M-estimation. \textit{The American Statistician}, 56(1), 29-38.}

\refmark{Wu, K. H. and Chen, L. P. (2025). A unified approach for estimating various treatment effects in causal inference. arXiv preprint \hyperlink{https://arxiv.org/abs/2503.22616}{arXiv:2503.22616}.}

\end{document}